\documentclass[11pt,a4paper]{article}
\pdfoutput=1

\usepackage{jheppub}
\usepackage{hyperref}
\usepackage{lscape}
\usepackage{authblk}
\usepackage{booktabs}
\usepackage{multirow}

\usepackage[normalem]{ulem}
\newcommand{\SU}{\ensuremath{\mathrm{SU}}}
\newcommand{\SO}{\ensuremath{\mathrm{SO}}}
\newcommand{\Uone}{\ensuremath{\mathrm{U}(1)}}

\title{Unification of Gauge Couplings in Radiative Neutrino Mass Models}

\collaboration{Claudia Hagedorn$^a$, Tommy Ohlsson$^b$, Stella Riad$^b$, and Michael A.~Schmidt$^c$}

\affiliation[a]{CP$^3$-Origins and Danish Institute for Advanced Study, University of Southern Denmark, Campusvej 55, 5230 Odense M, Denmark}
\affiliation[b]{Department of Theoretical Physics, School of Engineering Sciences, KTH Royal Institute of Technology, AlbaNova University Center, Roslagstullsbacken 21, 106 91 Stockholm, Sweden}
\affiliation[c]{ARC Centre of Excellence for Particle Physics at the Terascale,
School of Physics,\\ The University of Sydney, Physics Road, NSW 2006, Australia}

\emailAdd{hagedorn@cp3.sdu.dk}
\emailAdd{tohlsson@kth.se}
\emailAdd{sriad@kth.se}
\emailAdd{michael.schmidt@sydney.edu.au}

\abstract{
We investigate the possibility of gauge coupling unification in various radiative neutrino mass models, which generate neutrino masses at one- and/or two-loop level.
Renormalization group running of gauge couplings is performed analytically and numerically at one- and two-loop order, respectively.
We study three representative classes of radiative neutrino mass models:
(I) minimal ultraviolet completions of the dimension-7 $\Delta L=2$ operators which generate neutrino masses at one- and/or two-loop
level without and with dark matter candidates, 
(II) models with dark matter which lead to neutrino masses at one-loop level
and (III) models with particles in the adjoint representation of $\SU(3)$.
In class~(I), gauge couplings unify in a few models and adding dark matter amplifies the chances for unification.
In class~(II), about a quarter of the models admits gauge coupling unification.
In class~(III), none of the models leads to gauge coupling unification.
Regarding the scale of unification, we find values between $10^{14}$~GeV and $10^{16}$~GeV for models belonging to class~(I) without dark matter,
whereas models in class~(I) with dark matter as well as models of class~(II) prefer values in the range  $5 \cdot 10^{10}-5 \cdot 10^{14}$~GeV.
}
\arxivnumber{1605.03986}

\preprint{\begin{minipage}[t]{8cm}
		\begin{flushright}
CP3-Origins-2016-024 DNRF90\\
	DIAS-2016-024
\end{flushright}
\end{minipage}}

\begin{document}

\maketitle
\flushbottom

\section{Introduction}
\label{sec:introduction}

The Standard Model (SM) has been a great success in describing particle physics data.
However, it falls short in accommodating both massive neutrinos~\cite{Fukuda:1998mi,Ahmad:2002jz} and dark matter (DM)~\cite{Lundmark:1930,Zwicky:1933xx}, which indicate the existence of physics beyond the SM. 
Another, theoretical, indication for new physics is that the gauge couplings almost unify in the SM, which possibly points to the existence of a grand unified theory (GUT)~\cite{Georgi:1974sy}.
Without introducing new particles to the SM, neutrino masses can be encoded in the Weinberg operator~\cite{Weinberg:1979sa}. Their smallness forces the scale of this operator to be close to the GUT scale, 
assuming no suppression of its numerical coefficient. As is well-known, the
possible minimal ultraviolet (UV) completions at tree-level are the three types of the seesaw mechanism~\cite{Minkowski:1977sc,Yanagida:1980,Glashow:1979vf,Gell-Mann:1980vs,Mohapatra:1979ia,Magg:1980ut,Schechter:1980gr,Wetterich:1981bx,
Lazarides:1980nt,Mohapatra:1980yp,Cheng:1980qt,Foot:1988aq}, which can be naturally obtained within a GUT.

The smallness of neutrino masses can also be attributed to their generation at loop level. Such implementations are 
known as radiative neutrino mass models and usually predict new particles with masses within the reach of present and near-future experiments. The first radiative neutrino mass models were proposed at one-loop level in
Ref.~\cite{Zee:1980ai}, at two-loop level in Refs.~\cite{Cheng:1980qt,Zee:1985id,Babu:1988ki} and at three-loop level in
Ref.~\cite{Krauss:2002px}.\footnote{In all models discussed in the present study neutrinos are assumed to be Majorana particles. 
The first radiative neutrino mass model leading to Dirac neutrinos is found in
Ref.~\cite{Cheng:1977ir}.} The three-loop model in Ref.~\cite{Krauss:2002px} is
also the first radiative neutrino mass model, which predicts a DM candidate 
and thus also addresses the second shortcoming of the SM.

Given the appealing features of a GUT, it is highly interesting to study if it is, in fact, compatible
with radiative neutrino mass models. One of the essential
prerequisites for grand unification is unification of the gauge couplings, which we study in three representative classes of models.
Additional particles alter the renormalization group (RG) running of gauge couplings and thus can play an important role in achieving unification of the SM gauge couplings.
This is, in particular, true for the new particles needed for the generation of neutrino masses. 
This aspect has already been addressed in specific models. One example is a minimal realistic non-supersymmetric $\SU(5)$ GUT with scalars in the representations $15$ and $24$~\cite{Dorsner:2005ii,Dorsner:2005fq}, which requires an $\SU(2)$ triplet with TeV-scale mass as well as a light leptoquark in order to achieve gauge coupling unification as well as give rise to neutrino masses (via the type-II seesaw mechanism). 
 In Ref.~\cite{Arhrib:2009mz} an $\SU(5)$ GUT with fermions in the adjoint representation has been studied, which leads to neutrino masses via a combination of the type-I and type-III seesaw mechanisms and which requires an $\SU(2)$ triplet
 at the TeV-scale for gauge coupling unification. In Ref.~\cite{Dev:2015pga} the
	 authors have discussed gauge coupling
unification in the context of a left-right symmetric inverse
seesaw model~\cite{Mohapatra:1986bd} including a DM candidate and have shown that this model unifies to a non-supersymmetric $\SO(10)$ GUT.
Furthermore, gauge coupling unification has been discussed in a few radiative neutrino mass models. In Ref.~\cite{Sayre:2006ma}, a minimal trinification model with the gauge group $\SU(3) \times \SU(3)_L \times \SU(3)_R \times Z_3$~\cite{Achiman:1978rv,Glashow:1984gc} has been proposed.
In the context of an $\SO(10)$ model, a radiative seesaw mechanism~\cite{Ma:2006km} (the so-called scotogenic model) can be implemented, as shown in Ref.~\cite{Parida:2011wh}. 
Moreover, the possibility of gauge coupling unification has been discussed in
a TeV-scale $\SU(3) \times \SU(3)_L\times \Uone_X$
model~\cite{Singer:1980sw,Valle:1983dk}, where neutrino masses are generated via
loops of additional gauge bosons~\cite{Boucenna:2014dia}. In addition, an
$\SU(5)$ model in which neutrino masses are generated via the Zee mechanism has
recently been proposed \cite{Perez:2016qbo}. Beyond the
	RG running of gauge couplings in radiative neutrino mass models, there have been a few studies of
	the RG evolution of neutrino masses and leptonic mixing parameters: the first
	study of a one-loop radiative neutrino mass model~\cite{Bouchand:2012dx} explored the
	radiative seesaw model, while the first study of a two-loop
model~\cite{Babu:2014kca} focused on the Zee-Babu model.

In the present paper, we aim to perform a comprehensive study of gauge coupling unification in representative classes of radiative neutrino mass models instead of constraining ourselves to one specific model.
Thereby, we discuss RG running of the gauge couplings in all models analytically at one- as well as numerically at two-loop order.
We focus on models that are next-to-minimal compared to the three types of the seesaw mechanism~\cite{Minkowski:1977sc,Yanagida:1980,Glashow:1979vf,Gell-Mann:1980vs,Mohapatra:1979ia,Magg:1980ut,Schechter:1980gr,Wetterich:1981bx,
Lazarides:1980nt,Mohapatra:1980yp,Cheng:1980qt,Foot:1988aq}: 
UV completions of dimension-7 $\Delta L=2$ operators~\cite{Cai:2014kra}\footnote{These operators and similar ones up to dimension eleven 
have been listed in Refs.~\cite{Babu:2001ex,deGouvea:2007xp}. The construction of their minimal UV completions has been studied in Ref.~\cite{Angel:2012ug}.} and one-loop UV completions of the Weinberg operator~\cite{Restrepo:2013aga,Bonnet:2012kz}.
In addition, we consider models with particles in larger representations of $\SU(3)$~\cite{Angel:2013hla,FileviezPerez:2009ud}.
These classes of models have two characteristic features that are common to the majority of radiative neutrino mass models:
new colored particles and DM candidates in different $\SU(2)$ representations. The particles are assigned to representations with a dimension not larger than eight,
since an increase in dimension enhances the impact on the RG running~\cite{HALL198175,Machacek:1983tz},
causing a Landau pole (LP) in at least one of the gauge couplings. Furthermore, we do not discuss models with new massive vector bosons like the ones found in Refs.~\cite{Boucenna:2014dia,Deppisch:2016qqd}, since these require in general an extension of the gauge group of the model. 

The models of class~(I) consist of minimal UV completions of the dimension-7 $\Delta L=2$ operators introduced in Ref.~\cite{Cai:2014kra}.
Some of them have been discussed in more detail in Refs.~\cite{Zee:1980ai,Babu:2001ex,Babu:2009aq,Babu:2010vp,Babu:2011vb,He:2011hs}. In these models, neutrino masses are generated at one- and/or two-loop level and we consider the option of several generations of the new particles. Since these models rely on SM particles present in the loop(s), they relate neutrino masses to parameters of the SM, such as quark masses. 
We also consider the possibility of adding a DM candidate to these models, which is chosen among the representations discussed in Refs.~\cite{Cirelli:2005uq,Goodman:1984dc,Cirelli:2009uv}. Unification can be achieved in some models both with and without additional DM candidates. Without DM, the scale of gauge coupling unification is in the range $10^{14}-10^{16}$~GeV, whereas the range is lowered to $5 \cdot 10^{10}-5 \cdot 10^{14}$~GeV with DM.
Class~(II) contains the models discussed in Ref.~\cite{Restrepo:2013aga}. These are one-loop UV completions of the Weinberg operator using small $\SU(2)$ representations (up to dimension 3). They possess a DM candidate which is stabilized by a $Z_2$ symmetry. All such topologies have been presented in Ref.~\cite{Bonnet:2012kz}, with the most well-known model being the radiative seesaw model~\cite{Ma:2006km}.
All new particles are color neutral, and furthermore, the DM candidate must be electrically neutral.
In this case, gauge coupling unification can be achieved with the scale typically in the range $10^{13}-10^{14}$~GeV. Class~(III) comprises three models with at least one of the new particles in the adjoint representation of $\SU(3)$~\cite{Angel:2013hla,FileviezPerez:2009ud}. Proceeding like in the case of the models of class~(I), we consider the models of class~(III) without and with one type of DM candidate. In none of them, however, gauge coupling unification occurs. Hence, unless further modifications are made, they are not suitable as low energy models of a GUT.

The paper is organized as follows: In Sec.~\ref{sec:cond_uni}, we present the relevant RG equations (RGEs) for the SM gauge couplings at two-loop order, the conditions for gauge coupling unification which we impose, the approximations used in our numerical analysis and the analytic expressions following from gauge coupling unification at one-loop order.
Furthermore, we list the DM candidates that we add to the models in classes~(I) and (III).
In Sec.~\ref{sec:models}, we investigate if the gauge couplings unify in the various models analytically at one- and numerically at two-loop order. 
Based on our results we also discuss prospects for gauge coupling unification in other radiative neutrino mass models.
In Sec.~\ref{sec:disc}, we comment on corrections due to uncertainties in the gauge couplings at low energies and simplifications in our numerical analysis, the effects of the variation of the masses of the new particles, the differences in the RG running at one- and two-loop order, 
as well as the impact of Yukawa couplings on the RG running of the gauge couplings. In addition, we mention the RG running of the quartic scalar couplings and briefly discuss the embedding of the new particles of these models in representations of $\SU(5)$. Finally, in Sec.~\ref{sec:sc}, we summarize. In App.~\ref{app:betacoeff}, we list the contributions from particles in different representations to the one- and two-loop order coefficients $b_k$ and $b_{k\ell}$.

\section{Conditions for Unification}
\label{sec:cond_uni}

We study the RG running of the gauge couplings
\begin{equation}
\alpha_k = \frac{g_k^2}{4\pi} \,, \quad k=1,2,3 \, 
\end{equation}
in extensions of the SM with additional scalars and/or fermions responsible for the generation of neutrino masses at loop level.
The formulas for RG running at two-loop order 
are given in Ref.~\cite{Machacek:1983tz}. Assuming GUT normalization, i.e.~$g_1 =  \sqrt{5/3} \, g_Y$, the RGEs, with corresponding $\beta$-functions, for the gauge couplings are
\begin{equation}\label{eq:betafcn}
\frac{\mathrm{d}g_k}{\mathrm{d}t} = \beta_k(g)=b_k\frac{g_k^3}{(4\pi)^2}+\sum_{\ell=1}^3 \left[b_{k\ell}\frac{g_k^3g_\ell^2}{(4\pi)^4}+\frac{g_k^3}{(4\pi)^4} \, {\rm tr}(C^{u}_\ell Y_u^\dagger Y_u+C^{d}_\ell Y_d^\dagger Y_d+C^{e}_\ell Y_e^\dagger Y_e)\right] \,,
\end{equation}
where $t = \ln(\mu)$, $\mu$ being the energy scale.\footnote{The RGEs without contributions from the Yukawa couplings 
have been introduced in Ref.~\cite{HALL198175}. In addition, Eq.~\eqref{eq:betafcn} has to some extent already been discussed in Ref.~\cite{Fischler:1981is}.}
The one- and two-loop order coefficients, $b_k$ and $b_{k\ell}$, depend on the
particles in a given model.  
In general, these coefficients can be decomposed into two contributions, one coming from the SM, $b_k^{\rm SM}$ and $b_{k\ell}^{\rm SM}$, and another one from the additional particles in the model. Thus, the one- and two-loop order coefficients can be expressed as
\begin{equation}
b_k = b_k^{\rm SM}+\sum_{i=1}^N n_i \, b_k^i \,,\quad b_{k\ell} = b_{k\ell}^{\rm SM}+\sum_{i=1}^N n_i \, b_{k\ell}^i \,, 
\label{eq:bkbkl}
\end{equation}
where $N$ is the number of different representations, $i$, of new particles, $n_i$ the number of generations of particles in the representation $i$ and $b_k^i$ and $b_{k\ell}^i$ the contributions to the one- and two-loop order coefficients, respectively, due to 
the representation $i$. 
 The last term in the sum in Eq.~(\ref{eq:betafcn}) contains the contributions from the Yukawa couplings $Y_u$, $Y_d$ and $Y_e$ 
with the coefficients $C_k^u$, $C_k^d$ and $C_k^e$. Effects related to the Yukawa couplings of the new particles are neglected, as pointed out in Sec.~\ref{sec:Yukcoup}.
In the SM, the one- and two-loop order coefficients read~\cite{Jones:1981we}
\begin{equation}
(b_k^{\rm SM} ) 
= \begin{pmatrix}
	\frac{41}{10} \\[1mm]  -\frac{19}{6} \\[1mm] -7 
\end{pmatrix}
\,,\quad (b_{k\ell}^{\rm SM}) = 
\begin{pmatrix}
	\frac{199}{50} & \frac{27}{10} & \frac{44}{5}\\[1mm]
	\frac{9}{10} & \frac{35}{6} & 12\\[1mm]
\frac{11}{10} & \frac{9}{2} & -26 
\end{pmatrix}
\end{equation}
with $k, \ell=1,2,3$ and $C_k^f$ for $f=u, \, d, \,e$ are given by
\begin{equation}
	(C_k^u ) = \begin{pmatrix}
	-\frac{17}{10} \\[1mm] -\frac{3}{2} \\[1mm]  -2
\end{pmatrix} 
\,,\quad
	(C_k^{d} ) = \begin{pmatrix}
-\frac{1}{2} \\[1mm] -\frac{3}{2} \\[1mm] -2 
\end{pmatrix}
\,,\quad
	(C_k^{e} ) = \begin{pmatrix}
-\frac{3}{2} \\[1mm] -\frac{1}{2}  \\[1mm] 0
\end{pmatrix}
\end{equation}
with $k=1,2,3$.
We list the coefficients $b_k^i$ and $b_{k\ell}^i$  for all representations $i$, contained in the neutrino mass models discussed,  in App.~\ref{app:betacoeff}.

\subsection{Initial Conditions}
\label{sec:inicond}

The RG running of the gauge couplings $\alpha_1$, $\alpha_2$ and $\alpha_3$ is performed from the $Z$ mass, $M_Z=91.1876$~GeV, up to the scale of gauge coupling unification, $\Lambda$. The values of $\alpha_k$ at $M_Z$ are given by~\cite{Agashe:2014kda} 
\begin{equation}
\label{eq:initialnum}
\alpha_1 (M_Z) = 0.01704(1) \,, \quad
\alpha_2 (M_Z) = 0.03399(1)  \quad \mbox{and} \quad
\alpha_3 (M_Z) = 0.1185(6) \,.
\end{equation} 
The central value and error for $\alpha_3 (M_Z)$ are extracted directly from experiments, whereas those for $\alpha_1 (M_Z)$ and $\alpha_2 (M_Z)$ have been computed using 
the Weinberg angle and $M_Z$.\footnote{ 
The errors on $\alpha_1 (M_Z)$ and $\alpha_2 (M_Z)$ are obtained
by simply adding the (relative) errors of the Weinberg angle and $M_Z$ in quadrature.} 
In the following, the central values of $\alpha_k (M_Z)$ are used as initial conditions for the numerical analysis at two-loop order.

Within the SM (and using the simplification outlined in Sec.~\ref{sec:appthreshold}),
$\alpha_k (\Lambda_{\rm NP})$ at the mass scale $\Lambda_{\rm NP} $ of the new particles can be computed at two-loop order. Fixing $\Lambda_{\rm NP}=1$~TeV, as used throughout this analysis, and $\alpha_k (M_Z)$ 
to the central values in Eq.~(\ref{eq:initialnum}), we find
\begin{equation}
\label{eq:initialana}
\alpha_1 (\Lambda_{\rm NP}) = 0.01752 \,, \quad
\alpha_2 (\Lambda_{\rm NP}) = 0.03268 \quad \mbox{and} \quad
\alpha_3 (\Lambda_{\rm NP}) = 0.08972 \,.
\end{equation} 
In all analytical computations, these values are taken as initial conditions. 

\subsection{Unification}
\label{sec:unification}

We need to define the notion of gauge coupling unification in the present study in order to be able to classify the models.
The unification scale, $\Lambda$, is the scale at which the value $\alpha (\Lambda)$ of the three gauge couplings coincides. Graphically, this corresponds to the point where the three lines which describe the RG running of each of the gauge couplings intersect.
Due to numerical imprecision and several approximations made in our numerical analysis, it is not reasonable to require that all three gauge couplings are exactly the same at one single scale $\Lambda$.
Therefore, we allow for deviations in $\Lambda$ as well as in the corresponding values of $\alpha_k (\Lambda)$.
We consider the three values of the scale $\mu$ and $\alpha(\mu)$ given by the solutions to $\alpha_k (\mu)=\alpha_j (\mu)$, $k \neq j$, as vertices of a triangle. Then, we define $\Lambda$ and $\alpha (\Lambda)$ as the average of the three values 
of $\mu$ and $\alpha(\mu)$, respectively, at the vertices. 
Furthermore, we define the errors on $\log_{10}(\Lambda)$ and $\alpha(\Lambda)$ as the largest (in absolute value) of the differences of the values at the vertices of the triangle.\footnote{In the following, $\log_{10}(\Lambda)$ has to be read as $\log_{10}(\Lambda/{\rm GeV})$.} 
In our analysis, we consider gauge coupling unification to occur, if none of the two relative errors is larger than $8~\%$, i.e.  
\begin{equation}
	\frac{\Delta \log_{10}(\Lambda)}{\log_{10}(\Lambda)} \leq 8~\%  \quad \mbox{and} \quad
	\frac{\Delta \alpha^{-1}(\Lambda)}{\alpha^{-1}(\Lambda)} \leq 8~\% \,.
	\label{eq:delLambdaOverLambda}
\end{equation} 
In our analysis in Sec.~\ref{sec:models}, we observe that the relative error on $\alpha^{-1}(\Lambda)$ is approximately half of the error on $\log_{10}(\Lambda)$, and thus, the constraint on the former has virtually no effect. 

In addition, for each model we investigate whether an LP below the Planck scale, $M_{\rm Planck}$, is encountered or not. As definition of the scale $\Lambda^{\rm LP}_k$ of an LP in $\alpha_k$, we use $\alpha_k^{-1}(\Lambda^{\rm LP}_k)=1$.
 If
\begin{equation} 
\label{eq:LamLP}
 \Lambda^{\rm LP}_k \gtrsim 10^{3} \cdot \Lambda 
 \end{equation}
for all $\alpha_k$, the LP does not affect gauge coupling unification, in particular not $\Lambda$ itself, and therefore, such a model is considered viable. 
However, if this condition is not fulfilled for at least one $\alpha_k$, a more careful analysis is required.

\subsection{Approximations in the Numerics}

In order to simplify the numerical computations we make some approximations. In addition to assuming that all new particles have the same mass, $\Lambda_{\rm NP}=1$~TeV, we neglect all contributions from Yukawa couplings, except for the top Yukawa coupling, 
as well as all threshold effects.  

\subsubsection{Approximation for Yukawa Couplings}
\label{sec:Yukcoup}

The two-loop contributions from the Yukawa couplings in Eq.~\eqref{eq:betafcn} have, in principle, to be taken into account. 
However, their effect is usually small compared to other two-loop effects, and therefore, we do only consider the top Yukawa coupling and in general do
not evolve the Yukawa couplings in the present analysis. We make a conservative estimate of the contribution of the top Yukawa coupling, $y_t$, by setting $y_t=1$ 
in the RGEs for the gauge couplings.
In fact, this overestimates the effect of $y_t$, since it otherwise evolves to smaller values when the scale increases. 

The effects of the Yukawa couplings of the new particles are assumed to be negligible with respect to those of the top Yukawa coupling.
Such an assumption is reasonable given that these couplings are usually
constrained by flavor physics observables, if no special patterns among the
Yukawa couplings are imposed.

\subsubsection{Approximation for Threshold Effects}
\label{sec:appthreshold}

The RG running of the gauge couplings only depends on the particle content. At two-loop order, it is, however, necessary to consider model-dependent threshold effects. 
In the present analysis, we assume the full SM particle content at $M_Z$, and thus, threshold effects from the Higgs boson and the top quark are neglected. 
Furthermore, we do not take into account finite one-loop threshold effects arising from the new particles, since these can be absorbed in the re-definition of the matching scale~\cite{HALL198175,Weinberg198051}.

\subsection{Unification at One-Loop Order}
\label{sec:analytic-oneloop}

The RGEs for the gauge couplings $\alpha_k$ can be analytically solved at one-loop order
\begin{equation}
	\frac{1}{\alpha_k(\mu)}=\frac{1}{\alpha_k(\mu_0)}  -\frac{b_k}{2\pi}
	\ln\frac{\mu}{\mu_0} \,,
	\label{eq:RGE}
\end{equation}
where $\mu_0$ is the initial value. In this study, we set $\mu_0=\Lambda_{\rm NP}$. If we demand exact unification and restrict the unification scale to fulfill $\Lambda< M_{\rm Planck}$, we find
\begin{align}\label{eq:OneLoop}
	\sum_i n_i B^i_{kl}=
	\sum_i n_i \left(b_k^i - b_\ell^i\right) =
			\frac{2\pi}{L}
			\left[\alpha_{k,{\rm SM}}^{-1}(\Lambda)-\alpha_{\ell,{\rm SM}}^{-1}(\Lambda)\right] \,, \quad k,\ell =1,2,3 \;, \;\; k \neq \ell \,,
\end{align}
from equating the gauge couplings $\alpha_k(\Lambda)$ and $\alpha_\ell(\Lambda)$. 
We define 
\begin{equation}
\label{eq:BijkL}
B^i_{kl}= b^i_k-b^i_l \;\;\; ,  \;\;\; L = \ln \left( \frac{\Lambda}{\Lambda_\mathrm{NP}}\right)
\end{equation} 
and $\alpha_{k,{\rm SM}}(\Lambda)$ are the values of the gauge couplings at $\Lambda$ computed within the SM. 
In order to determine which particle content can give rise to unification, we use Eq.~\eqref{eq:OneLoop}. 
Unification requires equality of all three gauge couplings, which leads to two independent conditions of the form in Eq.~\eqref{eq:OneLoop}. 
For models with two types of new particles, e.g.~the models belonging to class~(I) without DM, we can solve for the number of generations $n_1$ and $n_2$ of the new particles in terms of $L$, namely
\begin{align}
n_1 & = \frac{2\pi}{L} \frac{B_{23}^2 \, \alpha_{1,{\rm SM}}^{-1}(\Lambda) + B_{31}^2 \, \alpha_{2,{\rm SM}}^{-1}(\Lambda) + B_{12}^2 \, \alpha_{3,{\rm SM}}^{-1}(\Lambda) }{ B_{23}^1 \, B_{31}^2 - B_{23}^2 \, B_{31}^1} \,, \label{eq:Sanalytic}\\ 
n_2 & = \frac{2\pi}{L} \frac{B_{23}^1 \, \alpha_{1,{\rm SM}}^{-1}(\Lambda)+
B_{31}^1 \, \alpha_{2,{\rm SM}}^{-1}(\Lambda) + B_{12}^1 \, \alpha_{3,{\rm
SM}}^{-1}(\Lambda) }{ B_{23}^2 \, B_{31}^1 - B_{23}^1 \, B_{31}^2 }\,. \label{eq:Sanalytic2}
\end{align}
Requiring the number of generations to fulfill $1 \leq n_i \leq 6$ and solving $n_i (\Lambda)$ for the lower ($n_i=1$) and upper ($n_i=6$) limits, we can determine the range of allowed values of $\Lambda$ for both $n_1$ and $n_2$.
The solutions correspond to one of the three different situations: (i) the ranges in $\Lambda$ for $n_1$ and $n_2$ overlap, (ii) the ranges do not overlap 
and (iii) the number of generations $n_1$ and/or $n_2$ is only equal or larger than 1 for $\Lambda$ larger than $M_{\rm Planck}$.  
Gauge coupling unification can only occur in situation~(i). Therefore, the models in situations~(ii) and (iii) can be discarded without further consideration. Hence, we focus the numerical analysis on models in situation~(i). 
In case there are more than two representations of new particles and/or the representations depend on some further parameter, e.g.~the dimension of the $\SU(2)$ representation, the unification conditions can still be determined for two of the unknown parameters.

\subsection{Dark Matter Candidates}
\label{sec:DM}

In the present study, we also consider the possibility to add DM candidates to the various models which do not contain one already, i.e.~to the models belonging to classes~(I) and (III). 
The choice of candidates is based on Refs.~\cite{Goodman:1984dc,Cirelli:2005uq,Cirelli:2009uv}. In order to fulfill the constraints from direct DM detection experiments, the candidates are restricted to fermions in representations with hypercharge $y=0$ as well as scalars
with either $y=0$ or $y=\frac 12$. Furthermore, from the RG running of the gauge couplings we can
impose an upper bound on the dimensionality of the $\SU(2)$ representation of the DM
candidate. This arises from the requirement that the RG running of the gauge couplings in the SM with only one generation of the DM candidate should remain
perturbative, i.e.~$\alpha_k(\mu) \lesssim 1$, at two-loop order up to $10^{16}$~GeV, the presumed scale of grand unification. The viable DM candidates are
listed in Tab.~\ref{tab:DM}. Note that we do not consider the possibility that the DM candidate is a singlet of the SM gauge group, since this does not affect the RG running. Such a DM candidate can, however, always be added to a model without changing our results.  
In order to achieve stability of the DM candidate, we assume an unbroken $Z_2$ symmetry to be present under which the DM candidate is odd and the rest of the particles of the model are even. 
This also implies that none of the scalars employed as DM can acquire a non-vanishing vacuum expectation value.

In the following, a particle $(r_3, r_2, y)$ is named according to its transformation properties under the SM gauge group $\SU(3) \times \SU(2) \times \Uone$, i.e.~it transforms in the representation $r_3$ under $\SU(3)$,
in $r_2$ under $\SU(2)$ and has hypercharge $y$. Furthermore, each particle carries
an index $S$ or $F$ indicating whether it is a scalar or a fermion.
\begin{table}[ht!]
\centering
\begin{tabular}{  c c c }
  \toprule			
  Scalar & Fermion & Scalar \\\midrule
  $(1,3,0)_S$ &  $(1,3,0)_F$ & $(1,2,\frac12)_S$\\[1mm] 
  $(1,5,0)_S$ &   $(1,5,0)_F$  & $(1,4,\frac12)_S$\\[1mm]  
  $(1,7,0)_S$  & &\\
   \bottomrule
\end{tabular} 
\begin{minipage}{12cm}
	\caption{Possible DM candidates that can be added to the radiative
	neutrino mass models belonging to classes~(I) and (III). We assume the DM particle(s) to be
stabilized by a $Z_2$ symmetry. The dimensionality of the $\SU(2)$
representation is limited by the requirement that gauge couplings remain
perturbative.}\label{tab:DM}
\end{minipage}
\end{table}

We do not perform a detailed study of DM phenomenology, since the discussion of gauge coupling unification is to a large extent independent of it. The DM annihilation cross section in general depends crucially on other couplings, which either do not enter the RGEs of the gauge couplings at two-loop order, like quartic scalar couplings, or whose contributions are negligible compared to the ones coming from the gauge couplings, like Yukawa couplings, as argued in Sec.~\ref{sec:Yukcoup}. Furthermore, the constraints on direct DM detection arising from interactions with gauge bosons are taken into account by an appropriate choice of the hypercharge of the DM candidate, see above. 
If the DM mass is larger than 1~TeV,
the cross section, relevant for indirect DM detection, strongly depends on the DM mass due to non-perturbative Sommerfeld enhancement~\cite{Hisano:2003ec}, whereas RG running of the gauge couplings only logarithmically depends on the particles' masses, see Eq.~(\ref{eq:RGE}).
Since we consider in most of the analysis particles with masses of 1~TeV or less, we refrain from performing a detailed study of this aspect of DM phenomenology. 

\section{Classes of Radiative Neutrino Mass Models}
\label{sec:models}

In this section, we study the RG running of gauge couplings in models, where neutrino masses are generated radiatively. 
We focus on classes of models that are next-to-minimal compared to the three types of the seesaw mechanism~\cite{Minkowski:1977sc,Yanagida:1980,Glashow:1979vf,Gell-Mann:1980vs,Mohapatra:1979ia,Magg:1980ut,Schechter:1980gr,Wetterich:1981bx,Lazarides:1980nt,Mohapatra:1980yp,Cheng:1980qt,Foot:1988aq} and that show two characteristic features of radiative neutrino mass models:
new colored particles and DM candidates in different $\SU(2)$ representations. 
We present our results for models belonging to the classes~(I) to~(III) introduced in Sec.~\ref{sec:introduction}.
First, in Sec.~\ref{sec:d7}, we analyze minimal UV completions of the dimension-7 $\Delta L=2$ operators with the possibility of several generations of the new particles. Then, in Sec.~\ref{sec:d7DM}, we consider the same type of models with only one generation of the new particles, but with a DM candidate (possibly in more than one generation). In Sec.~\ref{sec:MMDM}, we continue with class~(II), which consists of one-loop UV completions of the Weinberg operator containing a DM candidate. In Sec.~\ref{sec:octets}, we focus on the models of class~(III) which
contain fermions and/or scalars in the adjoint representation of $\SU(3)$.
Based on our results for models belonging to classes~(I) to~(III) we comment on the prospects for gauge coupling unification in other radiative neutrino mass models in Sec.~\ref{sec:othermodels}.

For each class, we perform both an analytic study at one-loop order and a numerical one at two-loop order. In the latter case, we compare our results to the ones from an approximate analytical two-loop study. 
The requirements for unification in the numerical and analytical studies are discussed in Secs.~\ref{sec:unification} and \ref{sec:analytic-oneloop}, respectively. 
These are used throughout unless otherwise stated. Based on the results at one-loop order, we can determine whether a numerical two-loop analysis is necessary or not.
For the numerical study, we use the software PyR@TE \cite{Lyonnet:2013dna} which computes the RGEs at two-loop order for a general gauge theory, specified by its gauge group, particle content and scalar potential. In the present analysis, 
we are only interested in the RG running of the gauge couplings. Hence, we only specify the gauge group (the one of the SM) and the particle content as input.

\subsection{Minimal UV Completions of Dimension-7 $\Delta L=2$ Operators}
\label{sec:d7}

In Ref.~\cite{Cai:2014kra}, 15 minimal UV completions of dimension-7 $\Delta
L=2$ operators with two new particles are found and possible ways to test them
at the LHC are discussed.
All models generate neutrino masses at one- and/or two-loop level.
The phenomenology of some of them has been studied in
detail, see Refs.~\cite{Babu:2009aq,Babu:2010vp,He:2011hs,Babu:2011vb,Zee:1980ai,Babu:2001ex}
for recent analyses.
The new particles are either two scalars 
or one scalar and one Dirac fermion, which give rise to three possible types of diagrams, see Figs.~1--3 in Ref.~\cite{Cai:2014kra}.
The particle contents for the 15 models are presented in
Tab.~\ref{tab:D7minimal} and labeled Si-j according to the diagram of type i.
New colored particles are one characteristic feature of these models and several of them are scalar leptoquarks, such as $(3,2,\frac{1}{6})_S$. These particles have a rich phenomenology \cite{Buchmuller:1986zs,Dorsner:2016wpm} 
and they have been recently used in order to explain neutrino masses together with several anomalies observed in the flavor sector, see e.~g.~Refs.~\cite{Pas:2015hca,Deppisch:2016qqd}. 
\begin{table}[hbtp]\centering\small
	\resizebox{\textwidth}{!}{
		\begin{tabular}{cccccccc}
		\toprule
		Model & S1-1 & S1-2 & S1-3 & S2-1 & S2-2 & S2-3 & S2-4 \\\midrule
		\multirow{2}{*}{Particles} &$(1,2,\frac{1}{2})_S$& $(3,2,\frac{1}{6})_S$ &
	$(3,2,\frac{1}{6})_S$   & $(1,2,-\frac{3}{2})_F$  &
		$(3,2,-\frac{5}{6})_F$  & $(3,1,\frac{2}{3})_F$  &
		$(3,1,\frac{2}{3})_F$  
\\[1mm]
& $(1,1,1)_S$ & $(3,1,-\frac{1}{3})_S$  & $(3,3,-\frac{1}{3})_S$ &
		$(1,1,1)_S$    & $(1,1,1)_S$&  $(1,1,1)_S$ &
$(3,2,\frac{1}{6})_S$ 
\\\midrule\midrule 
		S2-5 & S2-6 & S2-7 & S2-8 & S2-9 & S2-10 & S2-11 & S3  \\\midrule
  $(3,2,-\frac{5}{6})_F$  & $(3,2,-\frac{5}{6})_F$  & $(3,3,\frac{2}{3})_F$  &
$(3,2,\frac{7}{6})_F$  & $(3,1,-\frac{1}{3})_F$ & $(3,2,\frac{7}{6})_F$  &
$(1,2,-\frac{1}{2})_F$& $(1,3,-1)_F$  
\\[1mm]
 $(3,1,-\frac{1}{3})_S$ & $(3,3,-\frac{1}{3})_S$ & $(3,2,\frac{1}{6})_S$ &
$(1,1,1)_S$
			& $(1,1,1)_S$ & $(3,2,\frac{1}{6})_S$ & $(3,2,\frac{1}{6})_S$&
$(1,4,\frac{3}{2})_S$  
\\\bottomrule
\end{tabular}
}
\caption{Minimal UV completions of the dimension-7 $\Delta L=2$ operators in Tabs.~1-3 in Ref.~\cite{Cai:2014kra}.
In all models, two new particles are added in order to generate neutrino masses
at one- and/or two-loop level. }\label{tab:D7minimal}
\end{table}

\subsubsection{Analytic Results at One-Loop Order}
\label{sec:oneloop}

These models require two new particles, and thus, we can solve the conditions for unification given in Eq.~\eqref{eq:OneLoop} at one-loop order analytically. In model S1-2, following Eqs.~\eqref{eq:Sanalytic} and \eqref{eq:Sanalytic2}, we find the number of  particles required to achieve gauge coupling unification to be 
\begin{align}
\label{eq:1loopanaex1}
	n_1 & = \frac{\pi}{L}
	\left[-5\,\alpha_{1,{\rm SM}}^{-1}(\Lambda)+3\,\alpha_{2,{\rm SM}}^{-1}(\Lambda)+2\,\alpha_{3,{\rm SM}}^{-1}(\Lambda)\right] \,,\\
\label{eq:1loopanaex2}	
	n_2 & = \frac{\pi}{ L}
	\left[-5\,\alpha_{1,{\rm SM}}^{-1}(\Lambda)-9\,\alpha_{2,{\rm SM}}^{-1}(\Lambda)+14\,\alpha_{3,{\rm SM}}^{-1}(\Lambda)\right] \,.
\end{align}
Doing a similar computation for the other models Si-j, we find a finite overlap in $\Lambda$, i.e.~situation~(i), for models S1-2, S2-4, S2-6 and S2-11. Instead, for models S1-1, S1-3, S2-1, S2-7, S2-8, S2-10 and S3, we 
are in situation~(ii), whereas models S2-2, S2-3, S2-5 and S2-9 lead to situation~(iii). Hence, we only need to consider the four models in situation~(i) further. 
The overlap of the intervals in $\Lambda$ is the largest for model S1-2, whereas it is relatively small for the other three models S2-4, S2-6 and S2-11. 
In fact, in the numerical analysis we find two versions of model S1-2, where there is unification of gauge couplings, see Tab.~\ref{tab:2_uni}.

The one-loop analysis suggests the following ranges for the number of new particles: for model S1-2 there can be $3.4 - 5.5$ generations of particle $(3,2,\frac16)_S$ and up to six  of $(3,1,-\frac13)_S$, for model S2-4 we expect only one generation of particle $(3,1,\frac23)_F$ and six of $(3,2,\frac16)_S$, for model S2-6 we find $4.1 - 6$ generations of $(3,2,-\frac56)_F$ and only one generation of $(3,3,-\frac13)_S$, and finally, for model S2-11 we have to require only one generation of each of the new particles. Since two-loop effects might be significant, these numbers only give an indication of the actual number of generations. However, we find good agreement with the numerical results at two-loop order, displayed in Tab.~\ref{tab:2_uni},
except for the fact that model S2-6 does not lead to gauge coupling unification at two-loop order.

\subsubsection{Numerical Results at Two-Loop Order}

We present our results of the numerical two-loop analysis in Tab.~\ref{tab:2_uni}. 
We find four models that lead to unification of gauge couplings within the precision required: two versions of model S1-2, one version of model S2-4 as well as one version of model S2-11. 
Only for model S2-11, a single generation of each particle is sufficient for unification. In this model, the unification scale is $\Lambda \approx 1.2 \cdot 10^{14}$~GeV and thus lower than in the other three models.
In Fig.~\ref{fig:running4} (left panel), we show the RG running of the gauge couplings for model S1-2, with four generations of $(3,2,\frac{1}{6})_S$ and $(3,1,-\frac{1}{3})_S$, leading to gauge coupling unification at $1.8 \cdot 10^{16}$~GeV. 
This model has the highest scale of unification of the four models in Tab.~\ref{tab:2_uni}. The two-loop results are in agreement with the one-loop analysis in Sec.~\ref{sec:oneloop},
except for model S2-6. In this model, the particle content strongly affects the evolution of $\alpha_1$ and $\alpha_2$, and therefore, two-loop effects become relevant and gauge coupling unification cannot be obtained. 

\begin{table}[b!]\centering
\centering
\begin{tabular}{ cccccccc} 
  \toprule
  Model & P1 & P2 & $\Lambda$ (GeV) &
  $\alpha^{-1}(\Lambda)$ & 
  $\frac{\Delta \log_{10}(\Lambda)}{\log_{10}(\Lambda)}$ (\%) & $\frac{\Delta \alpha^{-1}}{\alpha^{-1}}$ (\%) \\
  \midrule
  S1-2 & $3\,(3,2,\frac{1}{6})_S$ & $(3,1,-\frac{1}{3})_S$ & $2.4\cdot 10^{15}$
  & 37.5 & 1.0 & 0.52 \\[1mm]
  S1-2 & $4\,(3,2,\frac{1}{6})_S$ & $4\,(3,1,-\frac{1}{3})_S$ & $1.8\cdot 10^{16}$ 
    & 35.1 & 1.6 & 0.80  \\[1mm]
  S2-4 & $(3,1,\frac{2}{3})_F$ & $5\,(3,2,\frac{1}{6})_S$ & $4.3 \cdot 10^{15}$ &
   32.3 & 1.8 &0.87 \\[1mm]
  S2-11 & $(1,2,-\frac{1}{2})_F$ & $(3,2,\frac{1}{6})_S$ & $1.2\cdot 10^{14}$ &
   38.4 & 1.2 & 0.61 \\
  \bottomrule
\end{tabular}
\caption{Models in Tab.~\ref{tab:D7minimal} where unification of gauge couplings is achieved with one to six generations of the two new particles. The number of generations needed is indicated in
front of the particle in representation $(r_3, r_2, y)_{S/F}$. In addition, we give the scale of unification $\Lambda$, the value of the gauge coupling $\alpha^{-1} (\Lambda)$ at this scale and the relative errors, quantifying the deviation from exact unification and fulfilling the constraints imposed in Eq.~(\ref{eq:delLambdaOverLambda}).}\label{tab:2_uni}
\end{table}
Finally, we also comment on the other models, which do not allow for gauge coupling unification.
Models S2-8 and S2-10 both reveal an LP at about $10^{17}$~GeV, whereas model S3 has an LP slightly below $10^{19}$~GeV. In models S2-3 and S2-9, unification cannot be achieved, since the RG running of the $\SU(2)$ gauge coupling cannot be affected with this particle content.
Similarly, in model S1-1, a sufficiently large effect on the RG running of
$\alpha_2$ can only be produced with more than six generations of the particle
$(1,2,\frac{1}{2})_S$. Furthermore, models S2-1, S2-2, S2-5, S2-8 and S2-10 have
particle contents which give rather large contributions to the RG running of
$\alpha_1$, and similarly, the particle contents of models
S1-3, S2-7 and S3 have a too large effect on the RG running of $\alpha_2$, preventing successful gauge coupling unification.

\subsection{Minimal UV Completions of Dimension-7 $\Delta L=2$ Operators with Dark Matter}
\label{sec:d7DM}

None of the new particles in the models Si-j can function as a (stable) DM candidate, compare Tabs.~\ref{tab:DM} and \ref{tab:D7minimal}. 
However, the DM problem can be addressed by the addition of one (or more) particles shown in Tab.~\ref{tab:DM} to any of the models Si-j, assuming that the DM candidate(s) is (are) odd under an additional $Z_2$ symmetry. 
The DM candidate(s) transform(s) non-trivially under the SM gauge group, and therefore, (it has) they have an impact on the RG running of the gauge couplings. 
For concreteness, we assume in the following models with one generation of each of the new particles, which participate in neutrino mass generation, and a single type of DM that can appear in several generations $n_d$, $1 \leq n_d \leq 6$. 

\subsubsection{Analytic Results at One-Loop Order}

We can distinguish two types of DM candidates with either $y=0$ or $y=\frac12$. In the first case, where $y=0$, the DM candidate only affects the RG running of $\alpha_2$, and therefore, the potential scale $\Lambda$ of gauge coupling unification is determined directly from the equality $\alpha_1 (\Lambda)=\alpha_3 (\Lambda)$. In general, the contribution from $n_d$ (real) scalars in the $\SU(2)$ representation of dimension $d$ to the one-loop order coefficient $b_2$ is given by 
\begin{equation}
\label{eq:b2dy0}
	b_{2}^{d} = \frac{1}{72} \, d \, (d^2-1) \, n_d \,,
\end{equation}
where we have used the Dynkin index $T(r)$, defined in Eq.~\eqref{eq:dynkin} in App.~\ref{app:betacoeff}. The contribution from a fermionic DM candidate in the same $\SU(2)$ representation as a real scalar is simply four times larger, compare Eq.~(\ref{eq:bscalarfermion}) in App.~\ref{app:betacoeff}. In the second case, where $y=\frac12$, the DM candidate affects the RG running of both $\alpha_1$ and $\alpha_2$. Hence, for $n_d$ generations of a (complex) scalar, transforming as $d$ under $\SU(2)$, the contributions to the one-loop order coefficients $b_1$ and $b_2$ read 
\begin{equation}
\label{eq:b2dy12}
	b_{1}^{d} = \frac{1}{20} \, d \, n_d \,, \quad b_{2}^{d} =  \frac{1}{36} \, d \, (d^2-1) \, n_d \,.
\end{equation}
Assuming again exact unification, we need to solve two equations, e.g.~$\alpha_1 (\Lambda) = \alpha_3 (\Lambda)$ and $\alpha_2 (\Lambda) = \alpha_3 (\Lambda)$, for two unknowns, e.g.~$\Lambda$ and $n_d$, which in turn depend on the third unknown $d$.

We can divide all models Si-j with DM into two categories depending on whether gauge coupling unification is possible or not. 
The models belonging to the latter category are models S1-3, S2-6, S2-7, S2-11 and S3, since in each case $n_d < 1$ (in several cases even negative) is required for exact unification. In particular, it is not possible to add further particles to model S2-11 without destroying the (already occurring) unification of gauge couplings.\footnote{As mentioned, we neither consider DM being a singlet of the SM gauge group nor do we consider additional particles to form complete representations of a larger gauge group like $\SU(5)$. In both cases, however, gauge coupling unification can be maintained in model S2-11 and, at the same time, DM is incorporated in this model.} 
 
The other models belong to the first category, where gauge coupling unification is possible in one or several cases. One example is model S1-2. Adding $n_d$ (real) $\SU(2)$ multiplets $(1,d,0)_S$, without specifying $d$ a priori, as DM candidates can lead to gauge coupling unification. Equating $\alpha_1 (\Lambda)$ and $\alpha_3 (\Lambda)$, which are not affected by the DM candidates, the scale $\Lambda$ is fixed, $\Lambda \approx 5.2 \cdot 10^{14}$~GeV, as well as the value of the gauge coupling at this scale, $\alpha^{-1}(\Lambda) \approx 39.0$. Then, the condition for unification of $\alpha_2$ with $\alpha_1$ (and thus also with $\alpha_3$) at $\Lambda$ can be recast in a condition on the contribution $b_2^d$ to the one-loop order coefficient $b_2$, namely
\begin{equation}\label{eq:b2}
	b_2^d=\frac{2\pi}{L}\left[ \alpha^{-1}_{2, {\mbox{\footnotesize S1-2}}}(\Lambda) - \alpha^{-1}_{1, {\mbox{\footnotesize S1-2}}}(\Lambda) \right] =  \frac{2\pi}{L}\left[ \alpha^{-1}_{2, {\rm SM}}(\Lambda) - \alpha^{-1}_{1, {\rm SM}}(\Lambda) \right] -\frac 25 \approx 0.699 \,,
\end{equation}
where $ \alpha_{k, {\mbox{\footnotesize S1-2}}}(\Lambda)$ denotes the value of the gauge coupling $\alpha_k$ in an extension of the SM with the particles of model S1-2, as given in Tab.~\ref{tab:D7minimal}.
Equating Eqs.~\eqref{eq:b2dy0} and \eqref{eq:b2}, we find for $d=3$ that $n_d \approx 2.1$ generations are necessary in order to achieve exact unification at one-loop order in model S1-2. 
Comparing this result to the numerical two-loop analysis we find good agreement regarding the scale $\Lambda$, the value of $\alpha^{-1} (\Lambda)$ as well as the number $n_d$ of generations.
This is also true for the other models Si-j and DM candidates with $y=\frac 12$, i.e.~the analytical results in general coincide with the outcome of the numerical study, summarized in Tab.~\ref{tab:2_DM}.

\subsubsection{Numerical Results at Two-Loop Order}
\label{sec:d7DMtwoloop}

We proceed to the numerical study, whose results are given in Tab.~\ref{tab:2_DM}.
As expected from the analytical results at one-loop order, adding DM to the models Si-j permits gauge coupling unification in several cases and in some occasions there is even more than one possibility for the DM type and number of generations. 
\begin{table}[b!]\centering
	\resizebox{\textwidth}{!}{
		\begin{tabular}{ c c c c c c c c}
  \toprule                       
  Model & P1 & P2 & DM & $\Lambda$ (GeV) & $\alpha^{-1}(\Lambda)$
  & $\frac{\Delta \log_{10}(\Lambda)}{\log_{10}(\Lambda)}$ (\%)
  & $\frac{\Delta \alpha^{-1}}{\alpha^{-1}}$ (\%) \\ \midrule
  S1-1 & $(1,2,\frac{1}{2})_S$ & $(1,1,1)_S$ & 2 $(1,3,0)_S$ & $7.7 \cdot 10^{13}$ & 39.6 & 1.2 & 0.61 \\[1mm]
  S1-1 & $(1,2,\frac{1}{2})_S$ & $(1,1,1)_S$ & 6 $(1,2,\frac{1}{2})_S$ &  $2.1\cdot10^{13}$ & 37.8 & 0.87 & 0.45 \\[1mm]
  S1-2 & $(3,2,\frac{1}{6})_S$ & $(3,1,-\frac{1}{3})_S$ & 2 $(1,3,0)_S$& $4.0\cdot 10^{14}$&38.8 & 3.1 & 1.6 \\[1mm]
  S1-2 & $(3,2,\frac{1}{6})_S$ & $(3,1,-\frac{1}{3})_S$ & 4 $(1,2,\frac{1}{2})_S$& $1.7 \cdot 10^{14}$ & 38.2 & 2.4 & 1.3\\[1mm]
  S2-1 & $(1,2,-\frac{3}{2})_F$ & $(1,1,1)_S$& $(1,5,0)_S$& $6.8\cdot 10^{10}$ & 31.4 & 1.0 & 0.53 \\[1mm]
  S2-1 & $(1,2,-\frac{3}{2})_F$ & $(1,1,1)_S$& $(1,4,\frac{1}{2})_S$& $5.9\cdot 10^{10}$ & 31.7 & 4.6 & 2.4 \\[1mm]
  S2-2 & $(3,2,-\frac{5}{6})_F$ & $(1,1,1)_S$& 3 $(1,3,0)_S$& $2.2\cdot 10^{12}$ & 30.5 & 0.67 &0.33\\[1mm]
  S2-2 & $(3,2,-\frac{5}{6})_F$ & $(1,1,1)_S$& 6 $(1,2,\frac{1}{2})_S$& $9.6 \cdot 10 ^{11}$ & 30.2 & 5.7 & 2.8 \\[1mm]
  S2-3 & $(3,1,\frac{2}{3})_F$ & $(1,1,1)_S$& $(1,5,0)_S$ & $3.9 \cdot 10^{13}$ & 35.4& 6.1 & 3.1\\[1mm]
  S2-3 & $(3,1,\frac{2}{3})_F$ & $(1,1,1)_S$& $(1,4,\frac{1}{2})_S$ & $2.8\cdot 10^{13}$ & 35.7 & 1.6 & 0.80 \\[1mm]
  S2-4 & $(3,1,\frac{2}{3})_F$ & $(3,2,\frac{1}{6})_S$& $(1,3,0)_F$  & $1.3\cdot 10^{14}$& 35.6 & 1.3 & 0.65  \\[1mm]
  S2-5 & $(3,2,-\frac{5}{6})_F$ & $(3,1,-\frac{1}{3})_S$ & 3 $(1,3,0)_S$ & $3.7\cdot 10^{12}$&30.4 & 0.46 & 0.22 \\[1mm]
  S2-8 & $(3,2,\frac{7}{6})_F$ & $(1,1,1)_S$ & $(1,5,0)_S$&$5.8\cdot 10^{10}$& 28.3 & 7.8 & 4.8 \\[1mm]
  S2-9 & $(3,1,-\frac{1}{3})_F$ & $(1,1,1)_S$ & $(1,3,0)_F$ & $2.6 \cdot 10^{14}$ & 37.9 & 0.54 & 0.28 \\[1mm]
  S2-10 & $(3,2,\frac{7}{6})_F$ & $(3,2,\frac{1}{6})_S$ & $(1,5,0)_S$ & $8.7\cdot 10^{10}$& 27.1 & 0.94 & 0.65 \\
  \bottomrule
\end{tabular}
}
\caption{Models in Tab.~\ref{tab:D7minimal} with additional DM that permit gauge coupling unification.
We assume throughout one generation of new particles, responsible for neutrino masses, and one type of DM with up to
six generations. The type of DM is chosen from Tab.~\ref{tab:DM}. We present the scale of unification $\Lambda$, the
value of the gauge coupling $\alpha^{-1}(\Lambda)$ at this scale and their relative errors.
}\label{tab:2_DM}
\end{table}

In general, we note that the types of DM, which are preferred, are those only modifying the RG running of $\alpha_2$. In addition, the dimension $d$ of the DM has to be smaller than seven and can only be $d=5$, 
if the DM is a scalar. Otherwise, the effect on the RG running of $\alpha_2$ would be too large, since this increases with the dimension $d$ of the representation, compare Eqs.~(\ref{eq:b2dy0}) and (\ref{eq:b2dy12}). 
Models where the added DM candidate has non-zero hypercharge have a lower scale of gauge coupling unification compared to models with DM with $y=0$, because any effect on the RG running of $\alpha_1$ forces $\Lambda$ to take smaller values. 
Given that the contribution to the RG running of $\alpha_k$ coming from a scalar is approximately four times smaller than the one from a fermion,\footnote{At one-loop order, this holds exactly, see Eq.~(\ref{eq:bscalarfermion}) in App.~\ref{app:betacoeff}, 
whereas this is no longer true at two-loop order. However, since two-loop effects are small, it is usually still true that one fermion has a very similar impact on gauge coupling running as four scalars.} 
it is clear that scalar DM is preferred. Thus, the most favorable types of DM are $(1,3,0)_{S/F}$ and $(1,5,0)_S$. The other types of DM allowed are $(1,4,\frac{1}{2})_S$ and $(1,2,\frac{1}{2})_S$, where the latter 
can be regarded as an additional Higgs doublet with vanishing vacuum expectation value.
Since DM particles must be color neutral, they can only contribute to the RG running of $\alpha_1$ and $\alpha_2$, which leads to $5 \cdot 10^{10}$~GeV$ <\Lambda < 5 \cdot 10^{14}$~GeV, i.e.~values of the gauge coupling unification scale $\Lambda$ 
that are about two orders of magnitude below $\Lambda$, obtained in models without DM, but with several generations of (possibly) colored particles.
For models S2-8 and S2-10 with an LP, the scale of unification given in Tab.~\ref{tab:2_DM} is at least three orders of magnitude smaller than the scale of the LP and thus the constraint in Eq.~(\ref{eq:LamLP}) is fulfilled.

In Fig.~\ref{fig:running4} (middle panel), we show the RG running of the gauge couplings for model S2-9 with DM in the representation $(1,3,0)_{F}$. In this case, a rather large unification scale is achieved, namely  
$\Lambda \approx 2.6 \cdot 10^{14}$~GeV, which is, as already mentioned, (still) nearly two orders of magnitude smaller than the highest scale of unification obtained for model S1-2 without DM, presented in Tab.~\ref{tab:2_uni}
and as well in Fig.~\ref{fig:running4} (left panel).

\subsection{Models with Dark Matter}
\label{sec:MMDM}

In Ref.~\cite{Restrepo:2013aga}, radiative neutrino mass models with a DM
candidate, which lead to neutrino masses at one-loop level, are discussed. These
models show a rich phenomenology which has been studied in numerous
publications, see e.g.~Refs.~\cite{Sierra:2008wj,Farzan:2010mr,Molinaro:2014lfa,Vicente:2014wga,Restrepo:2015ura,Longas:2015sxk,Arhrib:2015dez,Ibarra:2016dlb,vonderPahlen:2016cbw}.
We follow the convention in Ref.~\cite{Restrepo:2013aga}, where the models are
classified according to the four possible types of topologies of one-loop
diagrams, denoted T1-1, T1-2, T1-3 and T3. The topologies T1-i are box diagrams
that in general require four new particles, whereas the topology T3 is a
triangle diagram that requires at most three new particles. The new particles
are either scalars, Dirac or Majorana fermions. Furthermore, a $Z_2$ symmetry, which stabilizes the DM candidate(s), is assumed. Note that the $Z_2$ symmetry is unbroken, and therefore, the new scalars have vanishing vacuum expectation values.
In this type of models, all new particles are color neutral, which is necessary for a
viable DM candidate. Moreover, it is assumed that all new particles transform as singlets, doublets or triplets under $\SU(2)$. 
Each such possible choice leads to a model of certain topology T1-i or T3 that is enumerated by the type of topology followed by a capital Latin letter A, B, ... using the convention of Ref.~\cite{Restrepo:2013aga}.
The hypercharge of the new particles depends on a free parameter, denoted $m$ in the following and adjusted in such a way that at least one component of the multiplets of the new particles is electrically neutral.\footnote{Note that we use a different 
convention for the hypercharge than the authors of Ref.~\cite{Restrepo:2013aga}
 and thus values of the hypercharge given in the following have to be scaled by a factor of two in order to match the convention employed in Ref.~\cite{Restrepo:2013aga}.}
 Therefore, usually several choices of $m$ are viable for each model Ti-j-X. In addition, the value of $m$ indicates whether some of the new particles are in real representations or not. 
Finally, the number of models is reduced by the requirement that constraints from direct DM detection experiments are fulfilled. 

In total, we consider 35 (distinguishable) models with one generation for all new particles, which we investigate analytically and numerically.
If some of the new particles are in real representations of the SM gauge group and we study models with topologies T1-1, T1-3 and T3, a further version of the models can be
taken into account, in which two particles in the loop are identified with each other.
 An example is the radiative seesaw model~\cite{Ma:2006km}, that corresponds to model T3-B with $m=-1$, where 
 the two $\SU(2)$ doublet scalars $\phi\sim(1,2,\frac12)_S$ and $\phi^\prime\sim(1,2,-\frac12)_S$ are identified, i.e.~$\phi^\prime={\rm i} \sigma_2 \phi^\star$. 
However, in order to obtain two non-zero neutrino masses in such a model, two generations of the new particle coupling to the lepton doublets are required,\footnote{In Tab.~\ref{tab:Unification} this particle corresponds to the one denoted P3
for the three relevant topologies T1-1, T1-3 and T3.} e.g.~in the case of the radiative seesaw model at least two fermions being singlets of the SM gauge group are necessary.

\subsubsection{Analytic Results at One-Loop Order}
\label{sec:MMDMoneloop}

The free parameter $m$ only affects hypercharge, and thus, only the RG running of $\alpha_1$. In particular, the scale $\Lambda_{23}$, where $\alpha_2$ and $\alpha_3$
coincide, is independent of $m$.  
This simplifies the discussion and allows us to include possible deviations from exact unification, similarly to the numerical analysis, by imposing the constraints in Eq.~\eqref{eq:delLambdaOverLambda} in the analytical analysis.  
As explained in Sec.~\ref{sec:unification},  the scales $\Lambda_{ij}$, $i < j$, and the corresponding values of the gauge coupling $\alpha_i (\Lambda_{ij})$ can be viewed as the vertices of a triangle.
Since the constraint on the relative error on the scale is generally stronger than the one on the error on the value of the gauge couplings, we focus on the first inequality in Eq.~\eqref{eq:delLambdaOverLambda}, when deriving the range of values 
for $m$, which allows for gauge coupling unification at one-loop order in the models Ti-j-X. From the fact that $\alpha_1 (M_Z) < \alpha_2 (M_Z) <\alpha_3 (M_Z)$ and the knowledge that $\alpha_1 (\mu)$ increases with increasing $\mu$, whereas $\alpha_3(\mu)$ decreases, we find that either $\Lambda_{12} < \Lambda_{13} < \Lambda_{23}$ or $\Lambda_{23} < \Lambda_{13} < \Lambda_{12}$.
Hence, in both cases, the largest relative error on the scale arises from the difference in the logarithms of $\Lambda_{12}$ and $\Lambda_{23}$. Thus, we have
\begin{equation}
\label{eq:DelLamLamTijana}
\frac{\Delta \log_{10}(\Lambda)}{\log_{10}(\Lambda)} \approx \left| \frac{ \log_{10}\left(\Lambda_{12} \right)}{ \log_{10}\left(\Lambda_{23}\right)} -1 \right| \, ,
\end{equation}
where the actual unification scale $\Lambda$ is given by $\Lambda_{23}$ to good approximation.
  
As an example, we consider model T1-2-B assuming that all the new particles are in complex representations (i.e.~$m=0$ is not viable). The gauge couplings of $\SU(2)$ and $\SU(3)$ unify at $\Lambda_{23} \approx 9.0 \cdot 10^{12}$~GeV and $\alpha_2^{-1} (\Lambda_{23}) \approx 36.7$. 
Furthermore, we obtain
\begin{equation}
\label{eq:exeq}
\frac{\Delta \log_{10}(\Lambda)}{\log_{10}(\Lambda)} \approx \left|\frac{0.90-0.90 \, m-0.77 \, m^2}{7.4+1.2 \, m+m^2} \right| \,.
\end{equation}
Requiring exact unification, there are two solutions for $m$ which are $m \approx -1.8$ and $m \approx 0.64$. Allowing for deviations from exact unification, the expression in Eq.~(\ref{eq:exeq})
should be smaller than or equal to $8~\%$, which is fulfilled for $-2.2 \lesssim m \lesssim -1.4$ and $0.25 \lesssim m \lesssim 0.99$. Together with the constraints on the choice of $m$, see Ref.~\cite{Restrepo:2013aga}, we expect gauge coupling unification for $m=-2$ (with a relative error of $4.1~\%$). Furthermore, model T1-2-B for $m=2$ cannot lead to gauge coupling unification.
This expectation is indeed confirmed by the numerical analysis, carried out at two-loop order, see Tab.~\ref{tab:Unification}. Notice that also the values of the scale of unification and of the gauge coupling at this scale are in reasonable agreement between analytical and numerical study. 

As mentioned, depending on the model Ti-j-X and the choice of the parameter $m$ some of the particles are in real representations. These cases have to be treated separately,
since the contribution of such particles to the RG running of the gauge couplings is smaller by a factor of two compared to those in complex representations. 
Taking this into account, we find for model T1-2-B (with a gauge singlet fermion and an $\SU(2)$ triplet scalar being real) that $m$ has to lie in the interval $-2.2 \lesssim m \lesssim 0.68$ in order to achieve a relative error on the 
unification scale less than $8~\%$. In particular, the choice $m=0$ is expected to lead to gauge coupling unification at $\Lambda \approx 4.1 \cdot 10^{13}$~GeV with $\alpha^{-1}(\Lambda) \approx 38.4$.
Also these expectations agree well with the results of the numerical analysis, compare Tab.~\ref{tab:Unification}. 
 
Similarly, we study all other topologies and models, given in Ref.~\cite{Restrepo:2013aga}. It turns out that there is no further model in which the representations of all new particles are complex
and that permits gauge coupling unification. In the cases where some of the representations of the new particles are real, we find several models with gauge coupling unification, namely models T1-1-D with $m=\pm 1$, T1-1-G with $m=0$, T1-2-A with $m=0$, T1-3-A with $m=0$ and T3-A with $m=0$ and $m=-2$. Finally, we find no gauge coupling unification at one-loop order in cases with two of the new particles being identified and two generations of the new particle that couples to the lepton doublets. 
All in all, we find good agreement between these results and those of the numerical analysis.

\subsubsection{Numerical Results at Two-Loop Order}
\label{sec:MMDMtwoloop}

We proceed to the numerical analysis of the models.  
The models with gauge coupling unification are summarized in Tab.~\ref{tab:Unification}, where we list the particle content of each model, $\Lambda$, $\alpha^{-1}(\Lambda)$ and the relative errors on these two quantities.
\begin{table}[htbp]
\centering
\small
\resizebox{\textwidth}{!}{
	\begin{tabular}{c c c c c c c c c c}
  \toprule
  Model &$m$ &  P1 & P2 & P3 & P4 & $\Lambda$ (GeV) & $ \alpha^{-1}(\Lambda)$
  & $\frac{\Delta
  \log_{10}(\Lambda)}{\log_{10}(\Lambda)}$
  & $\frac{\Delta \alpha^{-1}}{\alpha^{-1}}$ \\ & & & & & & & & (\%) & (\%)\\ \midrule
  \multirow{2}{*}{T1-1-D} & 1 & $(1,2,\frac{1}{2})_S$ & $(1,1,0)_S$ &
  $(1,2,\frac{1}{2})_F$ & $(1,3,1)_S$ & $1.3\cdot 10^{13}$& 38.4 & 7.7&3.9
  \\[0.5mm]
			  & $-1$ & $(1,2,-\frac{1}{2})_S$ & $(1,1,-1)_S$ &
  $(1,2,-\frac{1}{2})_F$ & $(1,3,0)_S$ & $3.1\cdot 10^{13}$& 38.2 &
  3.2& 1.7  \\[3mm]
  T1-2-A & 0 & $(1,1,0)_F$ & $(1,2,\frac{1}{2})_S$ & $(1,1,0)_S$ &
  $(1,2,\frac{1}{2})_F$ & $5.3\cdot 10^{13}$& 39.4 & 4.1& 2.9 \\[3mm]
  \multirow{2}{*}{T1-2-B} & 0 & $(1,1,0)_F$ & $(1,2,\frac{1}{2})_S$ & $(1,3,0)_S$ & $(1,2,\frac{1}{2})_F$ & $4.6\cdot 10^{13}$ & 38.4 & 5.6 & 2.9\\[0.5mm]
			  & $-2$ & $(1,1,-1)_F$ & $(1,2,-\frac{1}{2})_S$ &
  $(1,3,-1)_S$ & $(1,2,-\frac{1}{2})_F$ &  $3.2\cdot 10^{12}$ & 35.9 &
  0.54 & 0.28 \\[3mm]
  T1-3-A & 0 & $(1,1,0)_F$ & $(1,2,\frac{1}{2})_F$ & $(1,1,0)_S$ &
  $(1,2,-\frac{1}{2})_F$ &$2.8\cdot 10^{13}$ & 37.7 & 6.5 & 3.3 \\[3mm]
  \multirow{2}{*}{T3-A} & 0 & $(1,1,0)_S$ & $(1,3,1)_S$ &  $(1,2,\frac{1}{2})_F$ &  - & $1.6\cdot 10^{13}$ & 37.3 & 4.4& 2.3  \\[0.5mm]
			& $-2$ & $(1,1,-1)_S$ & $(1,3,0)_S$ &
  $(1,2,-\frac{1}{2})_F$ & -& $4.0\cdot 10^{13}$ & 38.7 & 0.21 & 0.11 \\
\midrule
\midrule
  T1-3-A & 0 & $(1,1,0)_F$ & $(1,2,\frac{1}{2})_F$ & 2 $(1,1,0)_S$ & - & $6.9\cdot 10^{13}$ & 39.8 & 7.4 & 4.0 \\[3mm]
  T1-3-B & 0 & $(1,1,0)_F$ & $(1,2,\frac{1}{2})_F$ & 2 $(1,3,0)_S$ & - & $5.7\cdot 10^{13}$ & 38.9 & 2.5 & 1.3  \\
  \bottomrule
\end{tabular}
}
\caption{Models in Ref.~\cite{Restrepo:2013aga} that allow for gauge coupling unification.
In the last two rows, two particles are identified with each other (those transforming as $\SU(2)$ doublets) and two generations of particle P3 are needed.
 As in the tables in Secs.~\ref{sec:d7} and \ref{sec:d7DM}, the quantities $\Lambda$ and $\alpha^{-1}(\Lambda)$ are the unification scale and the value of the gauge coupling at this scale, respectively, and in the last two columns the relative errors
on these two quantities are given.
}\label{tab:Unification}
\end{table}

\begin{figure}[ht!]
\centering
\includegraphics[width=0.33\textwidth]{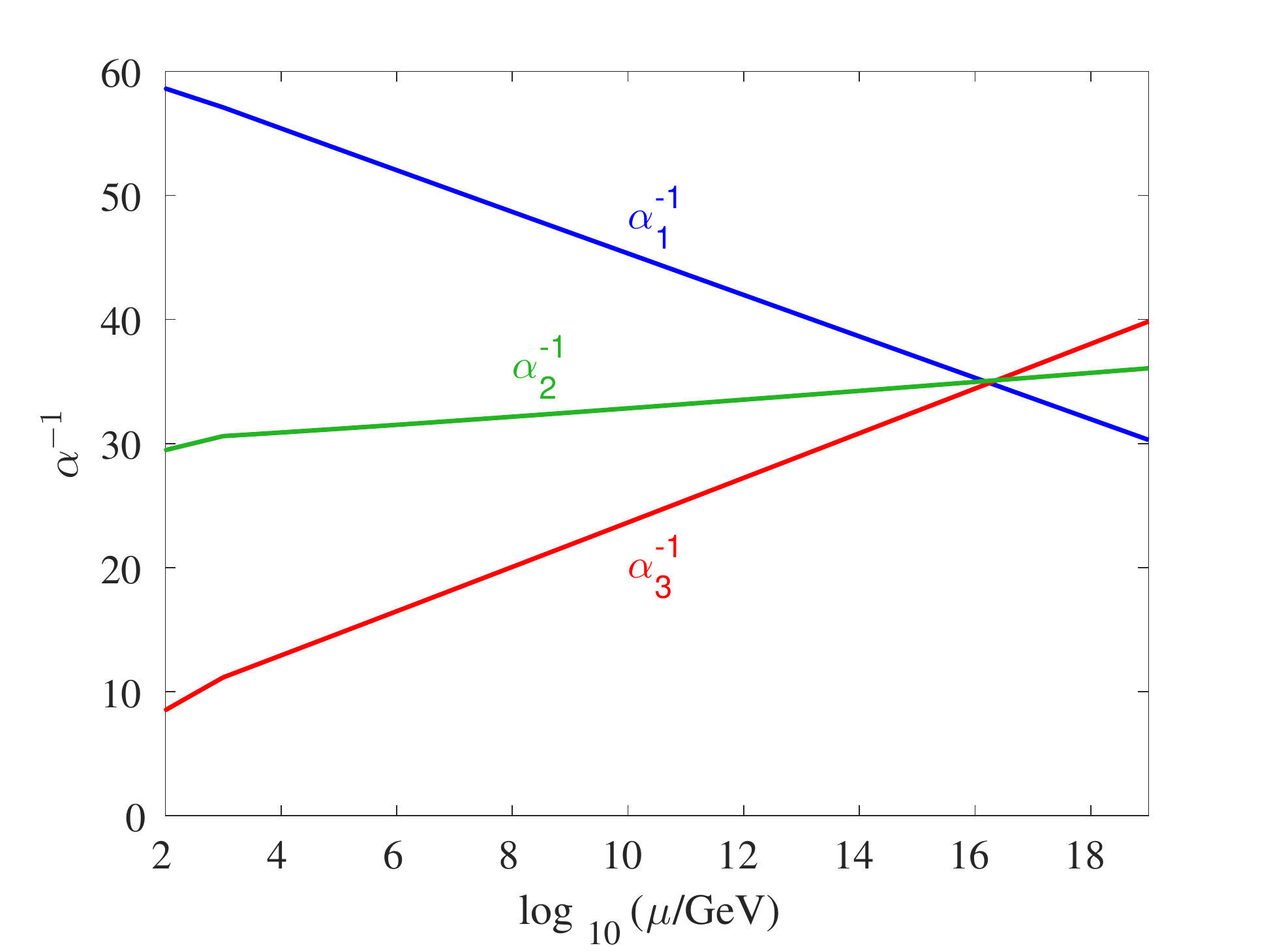}\hfill
\includegraphics[width=0.33\textwidth]{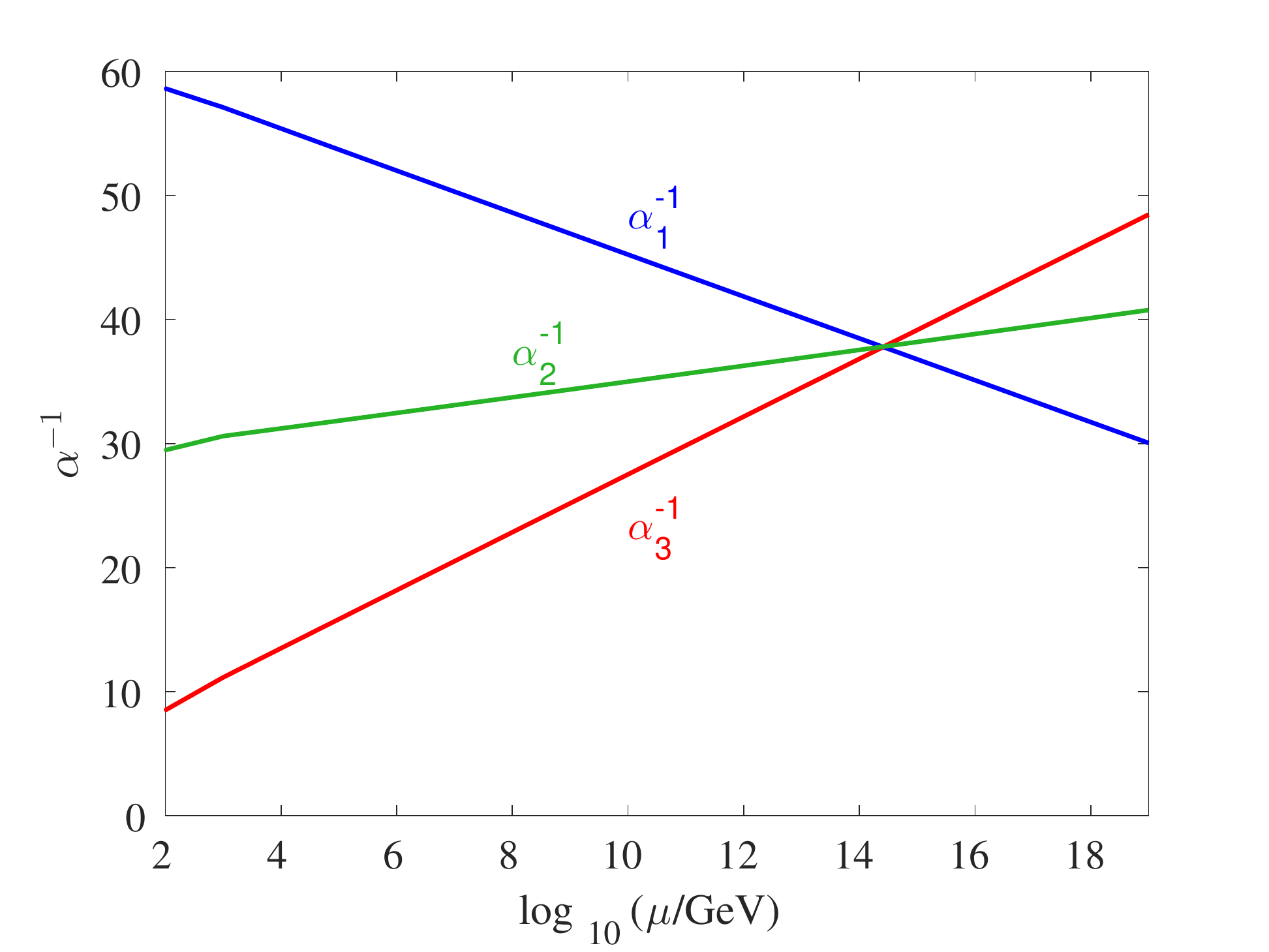}\hfill
\includegraphics[width=0.33\textwidth]{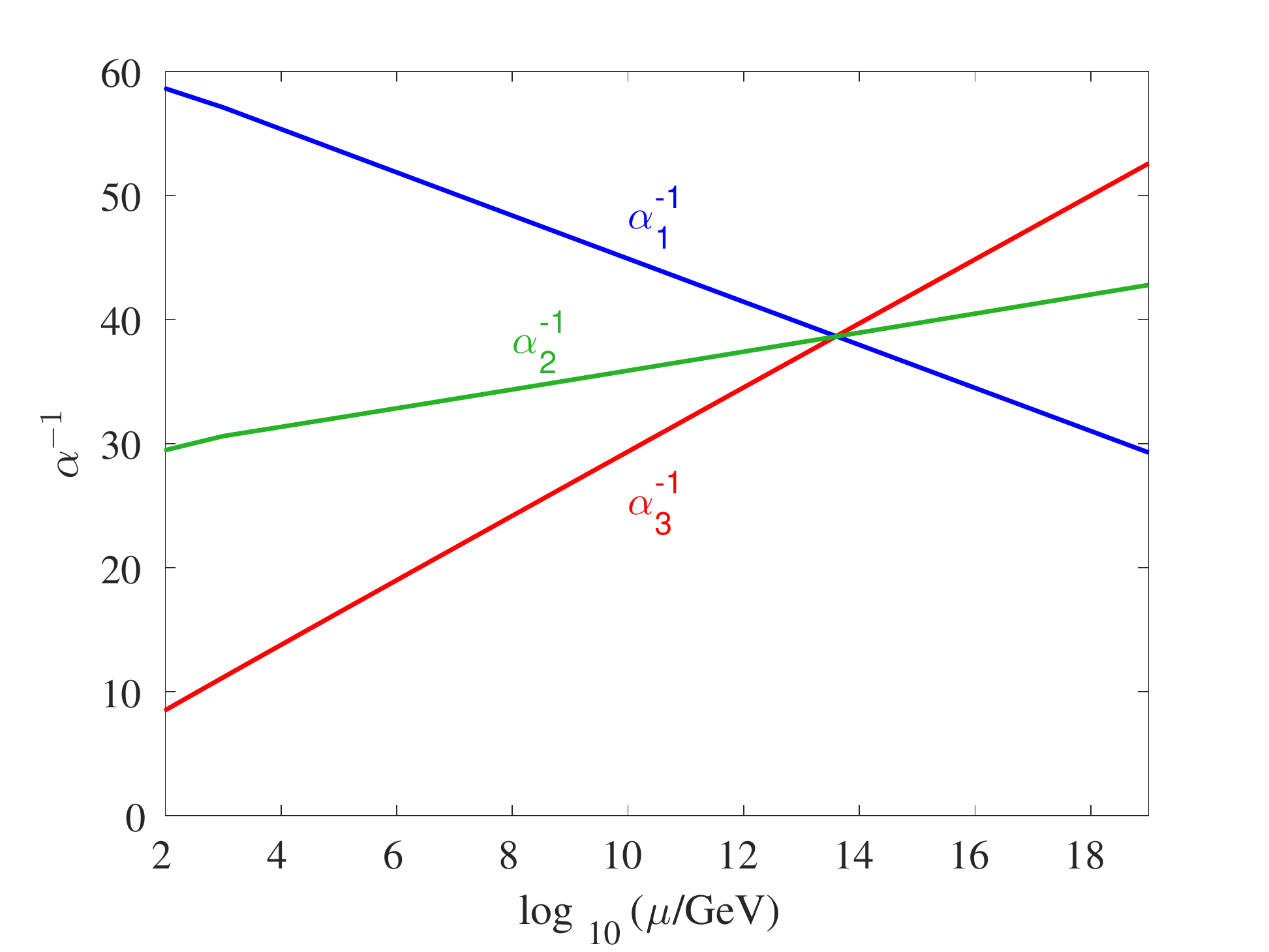} 
\caption{The RG running of the inverse of the gauge couplings $\alpha_i^{-1} (\mu)$ as function of the energy scale
$\mu$ for model S1-2 with particle content $4\,(3,2,\frac{1}{6})_S$ and
$4\,(3,1,-\frac{1}{3})_S$ (left panel), model S2-9 with DM $(1,3,0)_F$ (middle
panel) and model T3-A, $m = -2$ (right panel).
The gauge couplings $\alpha_1$, $\alpha_2$ and $\alpha_3$ for $\Uone$, $\SU(2)$ and $\SU(3)$ are shown in blue, green and red, respectively. Note that the kink at $\Lambda_{\rm NP}=1$~TeV is due to the new particles.}\label{fig:running4}
\end{figure}

Out of the 35 models, we find ten where gauge couplings unify at two-loop order with $\Lambda$ typically between $10^{13}$~GeV and $10^{14}$~GeV.
Thus, $\Lambda$ is usually below the scale of unification obtained in the models
discussed in Sec.~\ref{sec:d7}. This is mainly due to the fact that none of the new
particles in the models Ti-j-X carries color charge in contrast to most particles in models Si-j. 
 In Fig.~\ref{fig:running4} (right panel), we display the RG running of the gauge couplings in model T3-A for $m = -2$ as an example.
This model has been chosen, since the relative errors are the smallest among all models, see Tab.~\ref{tab:Unification}.
Note that in three of the ten models only SM gauge singlets and fermions and scalars in the $\SU(2)$ doublet representation are added, and thus, effectively, these models are like multi-Higgs doublet extensions of the SM regarding gauge coupling running. In particular, models T1-3-A with $m=0$ and two particles identified, T1-2-A with $m=0$ and T1-3-A with $m=0$ are (almost) equivalent to extensions of the SM with four, five and eight additional Higgs doublets, respectively.
Note also that in some of the models in which gauge coupling unification does not occur, 
$\alpha_1$ becomes non-perturbative at a scale below $M_{\rm Planck}$.

As mentioned, we find in general good agreement between the results of the analytical and the
numerical analysis. In the case of model T1-1-G with $m=0$, we observe that the
relative error on the unification scale is approximately $8~\%$ in the one-loop
analysis and small contributions at two-loop order eventually exclude this model.
Similarly, models with a relative error on the unification scale of
approximately $10~\%$ in the one-loop analysis are found to lead to gauge coupling unification, if studied numerically at
two-loop order. This is, in particular, true for models T1-3-A with $m=0$ and
T1-3-B with $m=0$, both with two particles identified.

%

\mathversion{bold}
\subsection{Models with Particles in the Adjoint Representation of $\SU(3)$}
\label{sec:octets}
\mathversion{normal}

There are a few models that generate neutrino masses at loop level where some of
the new particles, scalars and/or fermions, transform as the adjoint
representation of $\SU(3)$. We consider three of these models, presented in
Tab.~\ref{tab:octets}. In model U1, discussed in Ref.~\cite{Angel:2013hla} (see also for a detailed study of the phenomenology), neutrino masses
arise at two-loop level and two non-zero neutrino masses are
obtained, if two generations of the scalar are taken into account. Similarly, models U2
and U3, found in Ref.~\cite{FileviezPerez:2009ud}, lead to neutrino masses at
one-loop level and either two generations of the scalar or of the fermion are
necessary for two non-vanishing neutrino masses. The phenomenology of model U2 has been studied in
Refs.~\cite{Losada:2009yy,FileviezPerez:2010ch,Choubey:2012ux}.
\begin{table}[ht!]
\centering
\begin{tabular}{ccc}
  \toprule
  Model & Scalar & Fermion\\
  \midrule
   U1 & $(\overline{3},1,\frac{1}{3})_S$ & $(8,1,0)_F$ \\[1mm]
   U2 & $(8,2,\frac{1}{2})_S$ & $(8,1,0)_F$ \\[1mm]
   U3 & $(8,2,\frac{1}{2})_S$ & $(8,3,0)_F$ \\
    \bottomrule
\end{tabular} 
\begin{minipage}{12cm}
\caption{Models with particles in the adjoint representation of $\SU(3)$, where neutrino masses are generated radiatively. In model U1, neutrino masses arise at two-loop
level, see Ref.~\cite{Angel:2013hla}, while they are generated at one-loop level in models U2 and U3, see Ref.~\cite{FileviezPerez:2009ud}. 
In model U1, two generations of the scalar particle are necessary for achieving two non-zero neutrino masses, while in models U2 and U3, either two scalars or two fermions or both can be present.
}\label{tab:octets}
\end{minipage}
\end{table} 
As expected, the particles in the adjoint representation of $\SU(3)$ give rise to large contributions to the RG running of the gauge couplings. In some cases, these are so large that one of the gauge couplings becomes non-perturbative.
Indeed, only model U1 is free of LPs below $M_{\rm Planck}$. However, gauge coupling unification cannot be obtained in this case either. The conclusion is unchanged, if the two new particles are chosen to transform as the adjoint representation of $\SU(2)$ ($(\overline{3},3,\frac{1}{3})_S$ and $(8,3,0)_F$) instead of being singlets of $\SU(2)$. Rather, the situation is worsened, since in this
model, $\alpha_2$ assumes non-perturbative values at scales above $10^9$~GeV and $\SU(3)$ is not asymptotically free anymore.
Alternatively, we can add a DM candidate to model U1. However, this does not improve the RG running of the gauge couplings sufficiently in order to achieve their unification, independent of the type of DM considered and the number of generations of DM added.
In summary, we have not found a model with new particles in the adjoint representation of $\SU(3)$, which generates neutrino masses radiatively and, at the same time, allows gauge couplings to unify.

\subsection{Comments on Other Models}
\label{sec:othermodels}

In this section, we comment on the prospects for gauge coupling unification in radiative neutrino mass models that are not comprised in classes~(I) to (III). Among them are the
minimal UV completions of the higher-dimensional $\Delta L=2$ operators. For example,
in the case of dimension-9 $\Delta L=2$ operators with four fermions these completions mostly employ particles also 
needed in the case of dimension-7 $\Delta L=2$ operators~\cite{Angel:2012ug}, constituting class~(I) , e.g.~scalars transforming as $(1,1,1)$ under the SM gauge group. 
Thus, we might expect that gauge coupling unification can also be achieved in some of the models that lead to dimension-9 $\Delta L=2$  operators with four fermions. 
As discussed in Ref.~\cite{Angel:2012ug}, the minimal UV completions of dimension-9 $\Delta L=2$ operators with six fermions
 require in general new particles in different representations, e.g.~color sextets. In models with several particles in such a representation, gauge couplings are likely
 to become non-perturbative before unifying, since the impact of the six-dimensional representation of $\SU(3)$, $(\lambda_1, \lambda_2)=(2,0)$, is considerable on the RG running.
 In order to quantify this better, we compare its effect to the one of a particle in the adjoint representation of $\SU(3)$, $(\lambda_1, \lambda_2)=(1,1)$:
 the Dynkin index $T(r)$ of the six-dimensional representation is slightly smaller than in the case of the adjoint one (see Eq.~(\ref{eq:TrSU3}) in App.~\ref{app:betacoeff} for the definition of $T(r)$ in terms of the Dynkin labels $\lambda_1$
 and $\lambda_2$), while the eigenvalue $C_2 (r)$ of the quadratic Casimir operator is, indeed, slightly larger for the six-dimensional representation than for the adjoint one (see Eq.~(\ref{eq:C2rSU3}) in App.~\ref{app:betacoeff} for the definition of $C_2 (r)$
 in terms of the Dynkin labels $\lambda_1$ and $\lambda_2$). The Dynkin index $T(r)$ as well as $C_2(r)$ both determine the impact of a particle in a certain representation $r$ on the RG running of the gauge couplings, see e.g.~Eqs.~(\ref{eq:b3CS}) and ~(\ref{eq:Dckl}) in App.~\ref{app:betacoeff}. In addition, the six-dimensional representation of $\SU(3)$ is necessarily complex, whereas the adjoint one is real. Thus, we expect particles in the six-dimensional representation of $\SU(3)$
to have a similar or even larger impact on the RG running of the gauge couplings than the particles transforming as the adjoint one. Models with particles in the adjoint representation belong to class~(III) that we have studied. As shown in Sec.~\ref{sec:octets}, 
 gauge coupling unification does not occur in these models and in two of them at least one of the gauge couplings becomes non-perturbative below the presumable scale of unification.

A further type of models not included in classes~(I) to~(III) are
 two-loop UV completions of the Weinberg operator employing small $\SU(2)$ representations~\cite{Sierra:2014rxa}.
Yet, these have certain similarities to the models of class~(II)
	that represent one-loop UV completions of the Weinberg operator.\footnote{
 A noticeable difference between these two types of UV completions is the fact that the authors of Ref.~\cite{Sierra:2014rxa}
 do not require the existence of a DM candidate among the new particles, responsible for neutrino masses. However, 
 this only amplifies the possibilities of transformation properties of the new particles under the SM gauge group.} 
This allows us to comment also on the two-loop UV completions of the Weinberg operator. 
 In such models generally seven new particles are needed which are (at least) three more than in the case of the models of class~(II).\footnote{
 The number of new particles might be reduced, if identifications among them are possible, compare to models
 belonging to class~(II).} Thus, we expect larger effects on the RG running of the gauge couplings and an increased probability to
 encounter an LP as well as a tendency of having an even lower scale of gauge coupling unification.
 In Ref.~\cite{Sierra:2014rxa} all new particles are taken to be color neutral. However, it might be possible that some of them
 transform non-trivially under $\SU(3)$. This can be beneficial for raising the scale of gauge coupling unification, since then also
 the RG running of $\alpha_3$ is affected, see our findings for models of class~(I).

All models belonging to classes~(I) to~(III) generate neutrino masses at one- and/or two-loop level. However,
also several three-loop models have been discussed in the literature~\cite{Krauss:2002px,Ahriche:2014oda,Ahriche:2014cda,Chen:2014ska,Ahriche:2015wha,Okada:2015hia,Ahriche:2015loa}. 
Their realization usually entails that several of the new particles can be colored. In the case where some of the new particles are colored, we expect a result for gauge coupling unification similar to those obtained in the models of class~(I). In the case of only color neutral new states, which belong to $\SU(2)$ representations of larger dimension,\footnote{See Refs.~\cite{Cai:2016jrl,Ahriche:2016rgf} for one-loop models with particles in such representations.} the probability of encountering an LP in $\alpha_2$ and/or $\alpha_1$ increases, usually preventing successful gauge coupling unification at a sufficiently high energy scale.

In summary, the study of models belonging to classes~(I) to~(III) allows us -- to a certain extent -- to evaluate the prospects for gauge coupling unification in other types of models.

\section{Discussion}
\label{sec:disc}

We have shown that it is possible to achieve gauge coupling unification in several models that generate neutrino masses at one- and/or two-loop level, with and without DM. 
In this section, we discuss approximations and simplifications made in the analytical and numerical studies.
Furthermore, we comment on the RG running of the quartic scalar couplings, since it is known that such couplings can become non-perturbative below
a presumable gauge coupling unification scale.
Eventually, we address the question how these models can be embedded in an $\SU(5)$ GUT, since unification of gauge couplings is commonly taken as indication
for the existence of a larger gauge group than the one of the SM.  

\subsection{Sources of Corrections to Results}

Several corrections exist to the results shown in Sec.~\ref{sec:models}. In this section, 
we estimate, in particular, the size of the correction due to the uncertainty in the values $\alpha_i$ of the gauge couplings at low energies,
the effect of varying the mass scale $\Lambda_{\rm NP}$ of the new particles, the size of the two-loop contributions to the RG running of the gauge couplings
as well as the effect of the Yukawa couplings on the RG running of the gauge couplings. 

\mathversion{bold}
\subsubsection{Uncertainty in Values of Gauge Couplings at Low Energies}
\mathversion{normal}

There is an uncertainty in the measurements of the values of $\alpha_i (M_Z)$ at the $Z$ mass scale. It can be seen from Eq.~(\ref{eq:initialnum}) in Sec.~\ref{sec:inicond} that the relative uncertainties in $\alpha_1 (M_Z)$ and $\alpha_2 (M_Z)$ are very small compared to the one in $\alpha_3 (M_Z)$ and they can therefore be neglected. The uncertainty in $\alpha_3 (M_Z)$ amounts to about $0.5~\%$ at $M_Z$. 
In the case of model S2-11, the uncertainty in $\alpha_3 (M_Z)$ translates into a relative shift of $0.2~\%$ in $\log_{10}(\Lambda)$ with respect to the value mentioned in Tab.~\ref{tab:2_uni} in Sec.~\ref{sec:d7}, whereas we observe virtually no shift in the value of $\alpha^{-1} (\Lambda)$. We note that here and in the following this is the effect on the actual value of the unification scale $\Lambda$ and of the gauge coupling $\alpha^{-1}(\Lambda)$, respectively. In general, there is also an effect on the possibility of achieving gauge coupling unification, i.e.~an effect on $\Delta \log_{10}(\Lambda)$ and $\Delta \alpha^{-1}(\Lambda)$. However, the size of the latter is model dependent. The dominant effect of the uncertainty in $\alpha_3 (M_Z)$ is transmitted by the SM particles transforming under $\SU(3)$.  It can be slightly enhanced in models with one or more new particles charged under $\SU(3)$, since such representations directly affect the RG running of $\alpha_3$. In principle, only the effect due to the SM particles is visible in Fig.~\ref{fig:effects} (upper- and lower-left panels) in which we plot the results for model S2-11 and model T1-1-D with $m=-1$ (see Tab.~\ref{tab:Unification} in Sec.~\ref{sec:MMDM}). 

Assuming the full SM particle content at $M_Z$, especially the presence of the Higgs boson and the top quark, instead of taking them properly into account at their respective masses leads to a relative error on the gauge couplings, which is the largest for $\alpha_3$ and of the size $1.4~\%$ at $\Lambda_{\rm NP}$. The error on $\alpha_1$ and $\alpha_2$ is of the order $0.1~\%$ and can therefore be neglected. In Fig.~\ref{fig:effects} (upper- and lower-left panels), 
we show the effect of the combination of this effect together with the effect of the uncertainty in $\alpha_3 (M_Z)$ on the RG running of the gauge couplings in models S2-11 and T1-1-D with $m=-1$.

\subsubsection{New Particle Masses}

\begin{figure}[t!]
\centering
\includegraphics[width=0.33\textwidth]{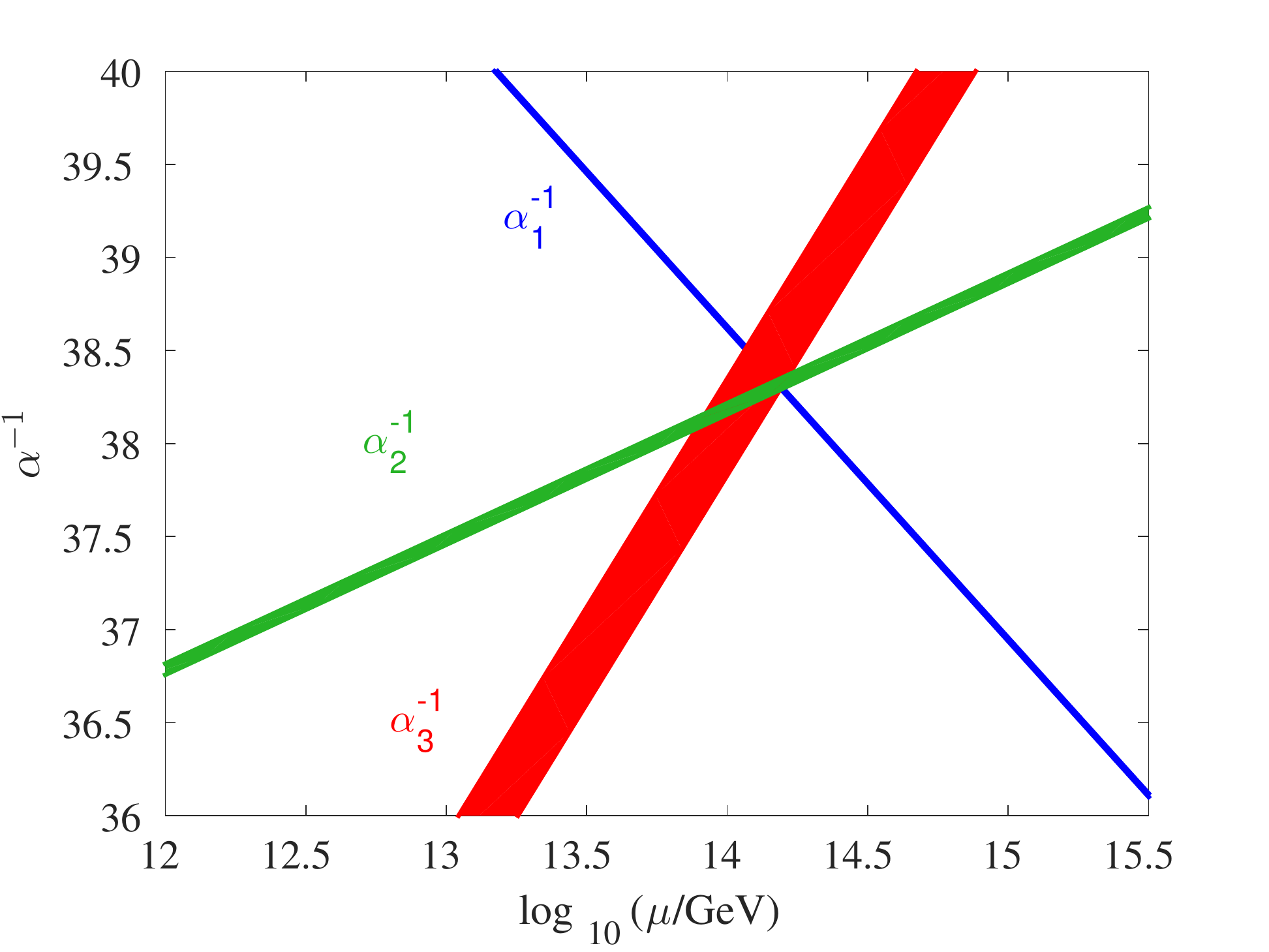}\hfill
\includegraphics[width=0.33\textwidth]{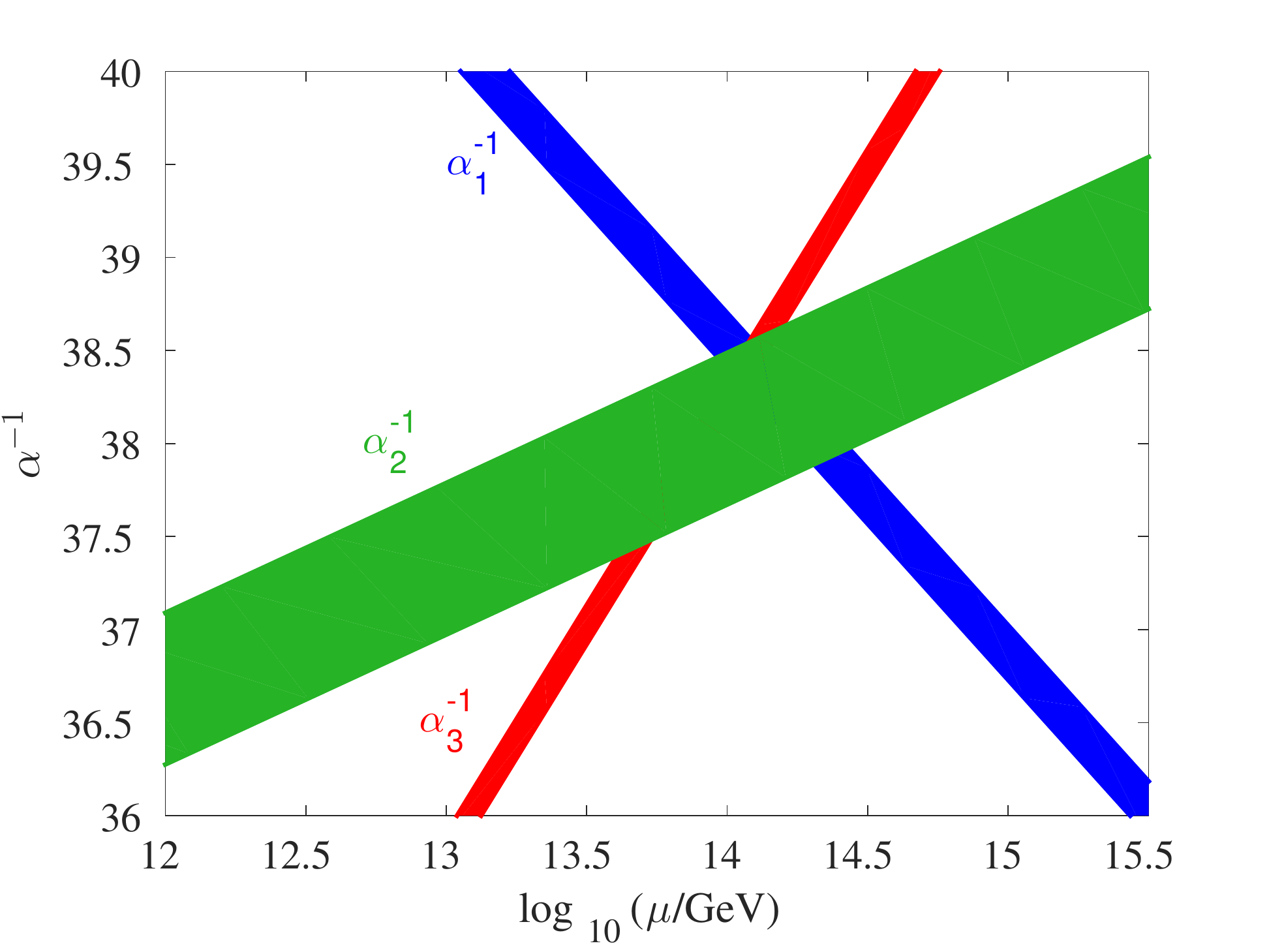}\hfill
\includegraphics[width=0.33\textwidth]{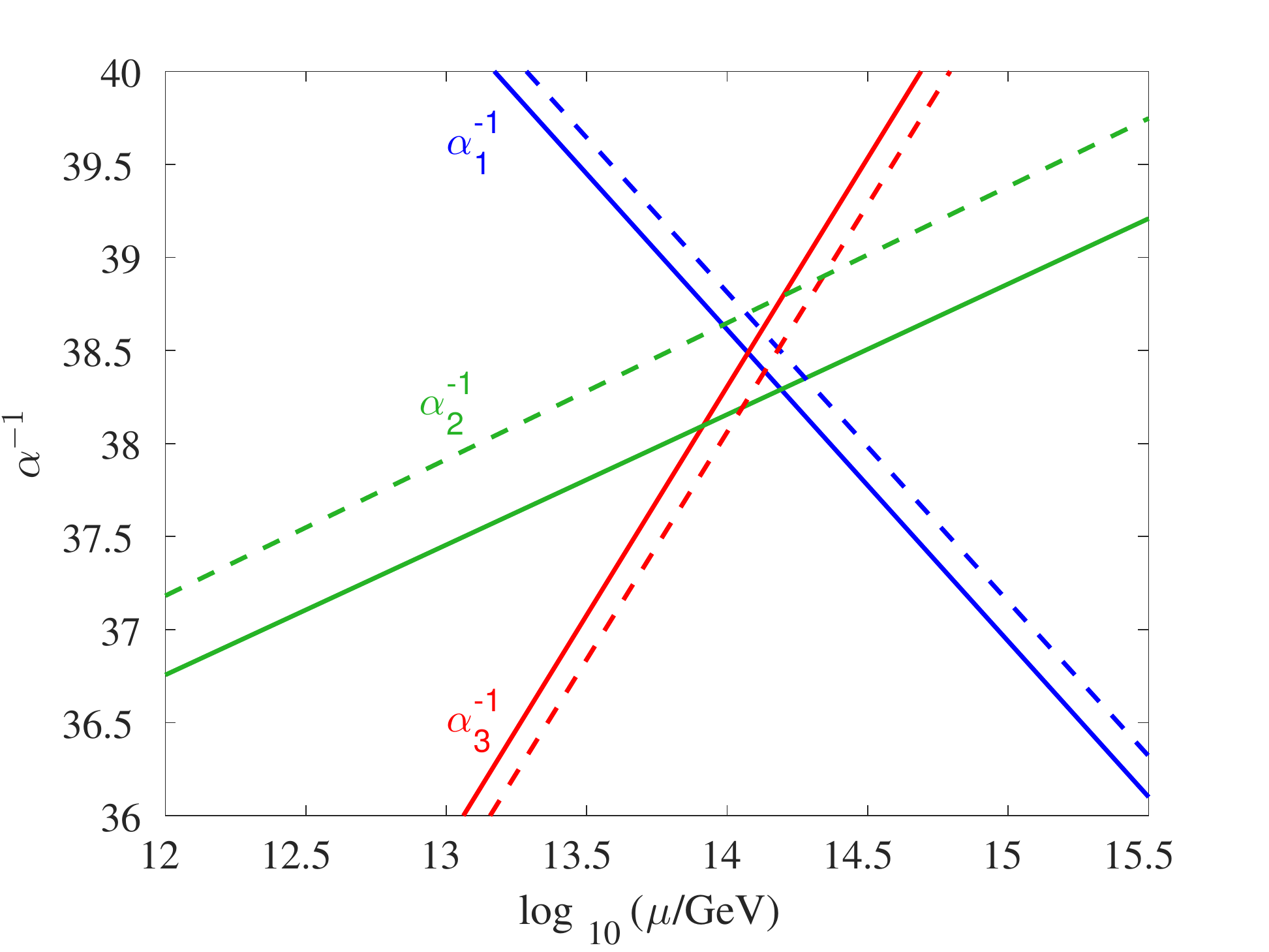} \\
\includegraphics[width=0.33\textwidth]{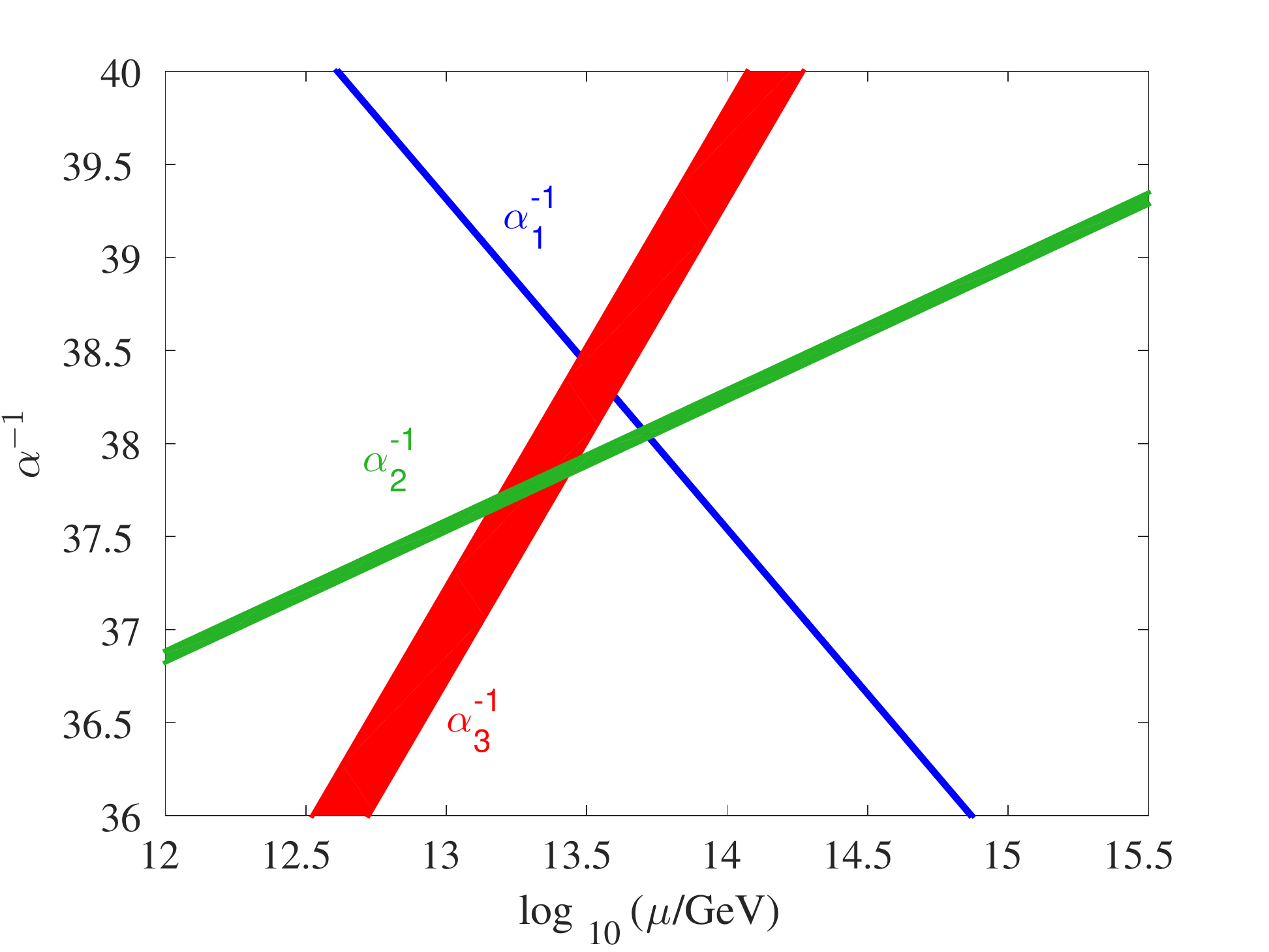}\hfill
\includegraphics[width=0.33\textwidth]{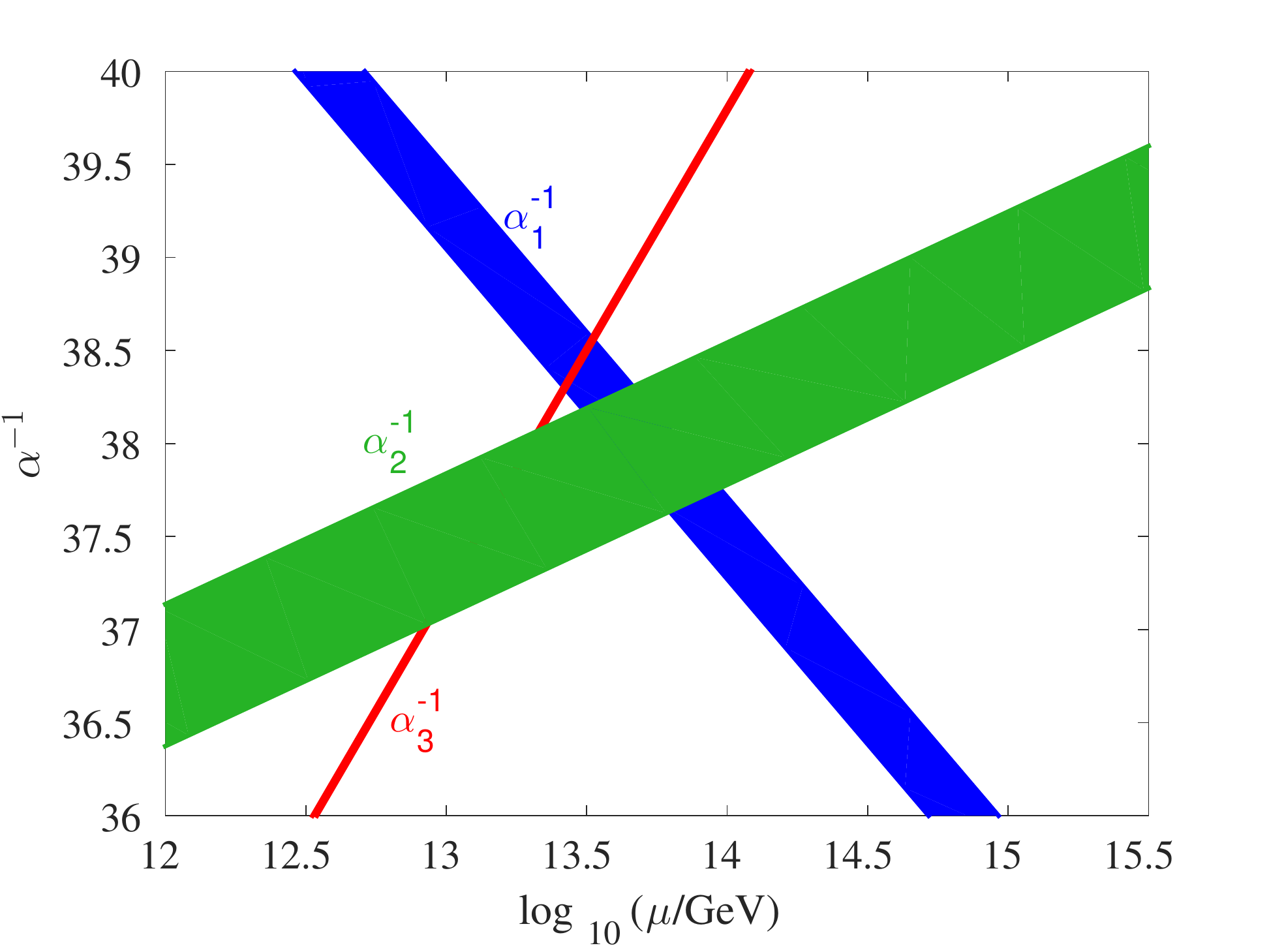}\hfill
\includegraphics[width=0.33\textwidth]{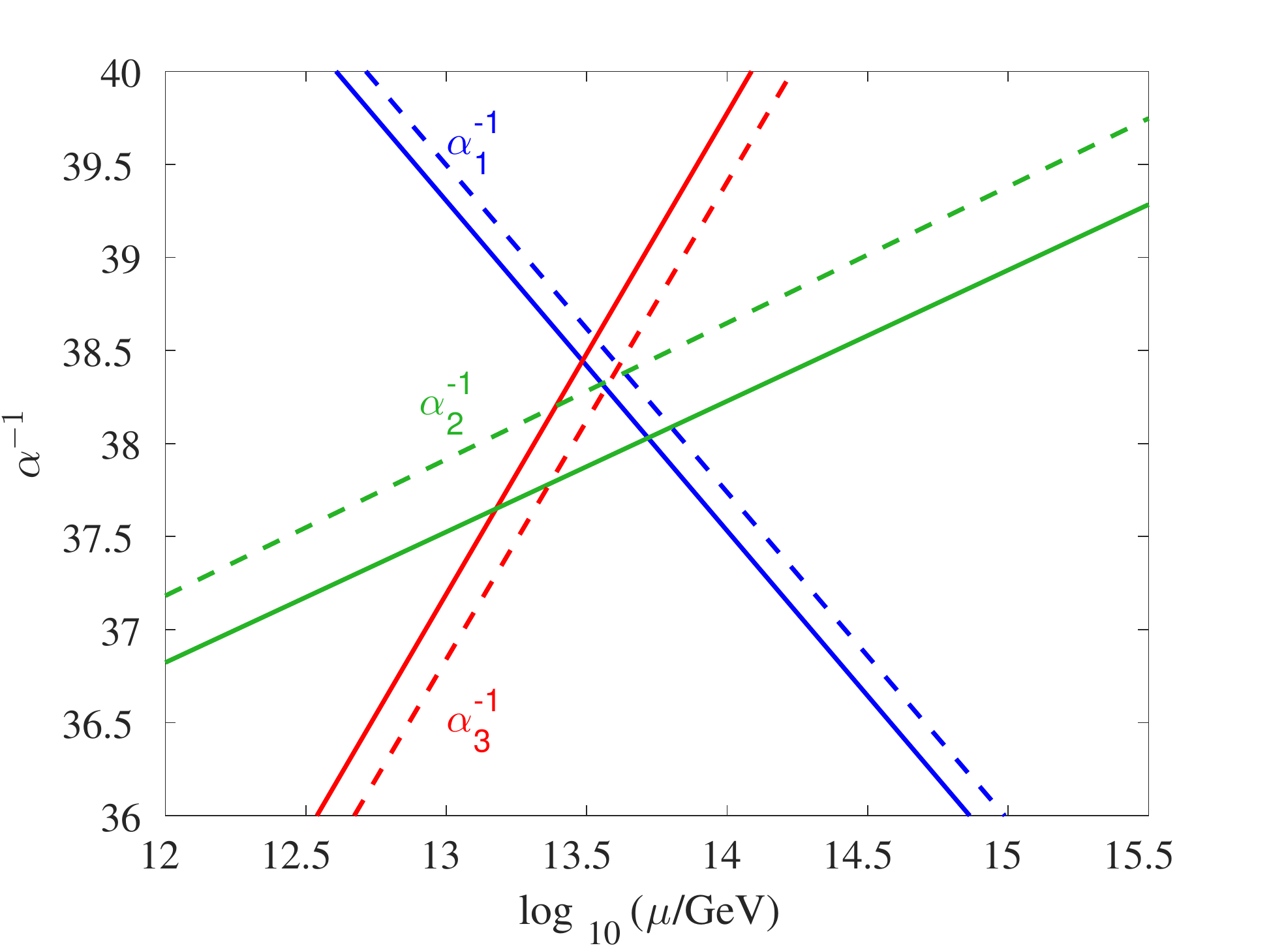}
\caption{The effect on the RG running of the gauge couplings $\alpha_i^{-1}(\mu)$ due to (a) the uncertainty in $\alpha_3 (M_Z)$ and the threshold effects in the SM due to Higgs boson and top quark for models S2-11 (upper-left panel) and T1-1-D, $m=-1$ (lower-left panel), 
(b) masses of the new particles ranging from 100 GeV to 5 TeV in models S2-11 (upper-middle panel) and T1-1-D, $m=-1$ (lower-middle panel) and
(c) two-loop contributions in models S2-11 (upper-right panel) and T1-1-D, $m=-1$ (lower-right panel). The gauge couplings $\alpha_1$, $\alpha_2$ and $\alpha_3$ for $\Uone$, $\SU(2)$ and $\SU(3)$ are shown in blue, green and red, respectively. 
Dashed lines display the RG running of $\alpha_i$ at one-loop order, while solid lines the RG running at two-loop order. }\label{fig:effects}
\end{figure}

Throughout the study we have made the simplifying assumption that the masses of all new particles are equal, $\Lambda_{\rm NP}=1$~TeV.
In this section, we estimate the effect of varying the masses between $100$~GeV
and $5$~TeV.
The mass range is chosen such that the masses 
allow for the possibility 
of detecting these particles, or at least some of them, at collider experiments. 
In Fig.~\ref{fig:effects} (upper- and lower-middle panels), we display this effect on the RG running in the two models S2-11 and T1-1-D with $m=-1$.
The boundaries of the bands in these plots are obtained for the two limiting cases, i.e.~if the masses of all new particles equal either the minimum value ($\Lambda_{\rm NP}=100$~GeV) or the maximum one ($\Lambda_{\rm NP}=5$~TeV).
 Clearly, the width of the bands depends on the dimensionality of the representations of the new particles and is larger for larger representations. In particular,
 we can see that the effect is the largest on $\alpha_2$, followed by $\alpha_1$, while $\alpha_3$ is hardly affected, in the two models.\footnote{Indeed, in model T1-1-D with $m=-1$, there is virtually no effect on $\alpha_3$, if $\Lambda_{\rm NP}$ is varied, simply because none of the new particles is charged under $\SU(3)$.} 
We can compute and compare the values of $\alpha^{-1}(\Lambda)$ and $\log_{10}(\Lambda)$ for the two limiting cases in which either all new particles have masses $\Lambda_{\rm NP}=100$~GeV
or all of them masses $\Lambda_{\rm NP}=5$~TeV. We find that the relative difference is about $2~\%$ for $\alpha^{-1}(\Lambda)$ and $0.4~\%$ for $\log_{10}(\Lambda)$ in the case of model S2-11. 
 Note that the values of $\alpha^{-1} (\Lambda)$ and $\log_{10}(\Lambda)$ for $\Lambda_{\rm NP}=1$~TeV, presented in Tab.~\ref{tab:2_uni}, lie in between those obtained in the two limiting cases. For values, obtained in model T1-1-D with $m=-1$, 
 similar statements hold.

Finally, we have also studied the possibility of raising the scale of gauge coupling unification
by assuming larger masses for the new particles. However, taking their masses to be $\Lambda_{\rm NP}=100$~TeV instead of $\Lambda_{\rm NP}=1$~TeV
still has only a small effect on the unification scale $\Lambda$. For example, in the case of model S2-11, the relative difference is about $3~\%$ for $\alpha^{-1}(\Lambda)$ and
 $0.5~\%$ for $\log_{10}(\Lambda)$ with respect to the values obtained for $\Lambda_{\rm NP}=1$~TeV. Thus, increasing $\Lambda_{\rm NP}$ does not sufficiently
help in raising the value of the unification scale.

%

\subsubsection{Two-Loop Contributions}
\label{sec:two-loop}

Since we have performed the analytical study at one-loop order, while we have studied the RG running of the gauge couplings numerically at two-loop order,
we also display examples of the effect of the two-loop contributions on the RG running. In Fig.~\ref{fig:effects} (upper- and lower-right panels), a comparison of the RG running at one- and two-loop order in models S2-11 and T1-1-D with $m=-1$ is made. 
Also the size of this effect depends in general on the number of new particles as well as their transformation properties under the SM gauge group. 
As an example, we discuss model S2-11. Comparing the values of $\alpha^{-1}(\Lambda)$ and $\log_{10}(\Lambda)$ extracted from the RG running of the gauge couplings at one- and two-loop order, respectively,
 we find that the relative difference in these values due to the different loop orders is about $1~\%$ for $\alpha^{-1}(\Lambda)$ and $0.6~\%$ for $\log_{10}(\Lambda)$. 

\subsubsection{Yukawa Couplings}

Yukawa couplings enter the RGEs of the gauge couplings at two-loop order, see Eq.~(\ref{eq:betafcn}). Following Sec.~\ref{sec:Yukcoup}, we neglect all Yukawa couplings apart from the top quark Yukawa coupling, which we fix to $y_t=1$ when evolving the gauge couplings between $M_Z$ and $M_{\rm Planck}$. In order to estimate the size of the effect resulting from Yukawa couplings of SM particles, we compare the value of $\Lambda$ for the cases with and without the (constant) top quark Yukawa coupling present in the $\beta$-functions of the gauge couplings. We find that the relative difference between the obtained values for $\log_{10}(\Lambda)$ is of the order of $0.1~\%$ in model S2-11, whereas the difference for $\alpha^{-1} (\Lambda)$ is completely negligible. 
The effects of the Yukawa couplings of the other SM particles as well as of the Yukawa couplings due to the new particles are neglected.

\subsection{RG running of Quartic Scalar Couplings}

Perturbativity of the gauge couplings does not guarantee perturbativity of all couplings in a model.
In particular, the quartic scalar couplings need attention. In
Ref.~\cite{Hamada:2015bra}, the RG running of especially the quartic self-couplings
of scalar particles, which transform under $\Uone$ and $\SU(2)$ but
not $\SU(3)$, is discussed. In their analysis, the authors assume that the new
scalars are added one at a time to the SM. In order to give a conservative
estimate of the scale of a possible LP, the initial conditions are that the
quartic self-couplings of the new scalars all vanish at a low energy scale
$M_X$. Furthermore, they put emphasis on the contribution from gauge boson loops
to the RG running of the couplings that is proportional to $\alpha_2^2$. The RG running of these couplings leads to an LP below $M_{\rm Planck}$ in several
cases, if $d$, the dimension of the $\SU(2)$ representation of the new scalars, is equal or larger than $4$,
$d\geq 4$, and the scale of the LP of the quartic scalar couplings (as well as
of the gauge couplings $\alpha_1$ and $\alpha_2$) is lowered, as $d$ and the
value of the hypercharge of the multiplet increase.

Thus, based on the results of Ref.~\cite{Hamada:2015bra} we can draw the
following conclusions for the present analysis: first, if the
new scalar is either in the representation $(1,2,\frac 12)_S$ or $(1,3,0)_S$,
the quartic scalar couplings remain perturbative up to $M_{\rm Planck}$ and thus
do not require further analysis; secondly, if scalars are added in the
representations $(1,4,\frac{1}{2})_S$, $(1,5,0)_S$ and $(1,7,0)_S$, e.g.~as
possible DM candidates, see Tab.~\ref{tab:DM}, the possible gauge coupling unification scale
should be compared with the scale of the LP of the quartic scalar couplings,
computed in Ref.~\cite{Hamada:2015bra}. Hence, for most of the models Si-j with
DM, see Sec.~\ref{sec:d7DM}, the LP occurs well above the unification scale, using the
conservative initial conditions of vanishing couplings at $M_X=\Lambda_{\rm
NP}=1$~TeV. The exception is model S2-3 in combination with a real scalar
quintuplet DM, $(1,5,0)_S$, where the scale of the LP is comparable to the gauge coupling unification scale, which is approximately $10^{13}$~GeV. 

The problem of encountering an LP of the quartic scalar couplings can be
mitigated by fermionic contributions to the $\beta$-function. This has been pointed
out in Ref.~\cite{Cai:2015kpa} and exemplified in the case of scalar minimal DM
transforming as $(1,7,0)_S$ under the SM gauge group.

\mathversion{bold}
\subsection{Embedding in an $\SU(5)$ Grand Unified Theory}
\mathversion{normal}

In case unification of gauge couplings is taken as indication for a gauge group larger than the one of the SM, e.g.~the grand unified group $\SU(5)$, to be present at high energies, the new particles responsible for 
radiatively generated neutrino masses and the DM candidates 
have to be embedded in representations of the GUT. Focusing for concreteness on $\SU(5)$, we observe that several of the new particles, e.g.~$(3,2,\frac 16)_S$ and $(3,1,\frac 23)_F$, can be straightforwardly embedded in 
small representations such as $5$, $10$ (or their conjugate representations) or $24$, since their quantum numbers coincide with the ones of an SM particle. 
The only representations, which correspond to some new particle contained in one of the models that can lead to gauge coupling unification according to our analysis and which need further consideration, 
are $(1,3,1)$, $(3,2,-\frac56)$, $(1,2,-\frac 32)$, $(3,2,\frac 76)$, $(1,4,\frac12)$ and $(1,5,0)$. 
Using Tab.~30 in Ref.~\cite{Slansky:1981yr}, we find that  
the representation $(1,3,1)$ can be embedded in $15$ of $\SU(5)$,
$(3,2,-\frac56)$ in $24$, $(1,2,-\frac 32)$ in $40$, $(3,2,\frac 76)$ in
$\overline{45}$ and $(1,4,\frac12)$ in $70$. In addition, we have explicitly checked that the representation $(1,5,0)$ can be embedded in $200$ of $\SU(5)$ by making use of the generating function given in Ref.~\cite{Patera:1981pn}.

\section{Summary}
\label{sec:sc}

We have investigated the RG running of the SM gauge couplings and their possible unification for three classes of neutrino mass models where neutrino masses are generated radiatively at one- and/or two-loop level.  Class~(I) consists of minimal UV completions of the dimension-7 $\Delta L=2$ operators. These models, denoted Si-j, contain two types of new particles, in addition to those of the SM. In our study, these new particles either appear in several generations or 
in one generation together with (several copies of) one type of DM. Class~(II) consists of the models Ti-j-X, which are one-loop UV completions of the Weinberg operator containing, in general, three or four new particles. Assuming that these new particles
are odd under an additional $Z_2$ symmetry (the SM particles are even), one of them constitutes a DM candidate. 
Class~(III) contains three models Ui with at least one of the new particles in the adjoint representation of $\SU(3)$.

The results of our study can be used as a guideline for building models of grand unification in which neutrino masses
are generated radiatively. The main points are:
(i) Gauge coupling unification in one step can be achieved in models belonging to classes~(I) and (II). 
However, the scale of gauge coupling unification is generally lower in models of class (II) compared to those of class~(I). This
shows that obtaining a unification scale high enough in order to satisfy limits derived from the absence of proton decay
requires the presence of new colored states with appropriately chosen values of the hypercharge in order not to affect too much
the RG running of $\alpha_1$.  (ii) Adding a DM candidate, which belongs to an SU(2) multiplet, to a model in order
to achieve gauge coupling unification impacts on the RG running of $\alpha_2$ (and possibly $\alpha_1$), while $\alpha_3$ is necessarily unaffected.
Hence, the scale of gauge coupling unification tends to be in general lower than in the corresponding model without DM candidate.

	The details of the results of the analytical and numerical studies, performed at one- and
two-loop order, respectively, can be summarized as follows:
a few of the models belonging to class~(I) allow for unification of the gauge couplings, if the new particles can appear in several generations. Since these new particles usually transform non-trivially under $\SU(3)$, the unification scale lies between $10^{14}$ and $10^{16}$~GeV.
Two examples of this type of models are given in Tab.~\ref{tab:conclusions}: the first one has the highest value of the unification scale $\Lambda$ which is achieved with four generations of each new particle. In the second one, only two new particles are added to the SM, which, however, results in a rather low value of $\Lambda$. If instead only one generation of each new particle is assumed and (several copies of) one type of DM is (are) present, we find more than ten possible
models in which the gauge couplings unify at some high energy scale. The
preferred range of this scale is $5 \cdot 10^{10}~\text{GeV} \lesssim \Lambda \lesssim 5
\cdot 10^{14}~\text{GeV}$. Again, one example is listed in
Tab.~\ref{tab:conclusions}. In about a quarter of the models of class~(II), we
find unification of the gauge couplings and in this case the unification scale
tends to be rather low, of the order of $10^{13}$~GeV, see the example in
Tab.~\ref{tab:conclusions}. In addition, depending on the choice of the
parameter $m$, which parameterizes the hypercharge of the new particles, the
number of types of new particles can
be reduced (two particles can be identified, while two copies of the new particle coupling to the lepton doublets must exist) and two additional models with gauge coupling unification are found. This time the unification scale is slightly larger. 
Finally, none of the models in class~(III) (taking into account the possibility of having more
than one generation of the new particles as well as adding one type of DM to the model) allows for unification of the gauge couplings. 
Based on these results, we have also commented on the prospects for gauge coupling unification in models of radiative neutrino mass generation that are not comprised in classes~(I) to (III).
\begin{table}[ht!]\footnotesize
\centering
\resizebox{\textwidth}{!}{
	\begin{tabular}{c c c c c c c c }
  \toprule			  
  Model & P1 & P2 & P3  & $\Lambda$ (GeV) &  $\alpha^{-1}(\Lambda)$ & $\frac{\Delta
  \log_{10}(\Lambda)}{\log_{10}(\Lambda)}$& $\frac{\Delta\alpha^{-1}}{\alpha^{-1}}$ \\  & & & & & & (\%) & (\%)\\
  \midrule
 S1-2 & $4\,(3,2,\frac{1}{6})_S$ & $4\,(3,1,-\frac{1}{3})_S$  & - & $1.8\cdot 10^{16}$ 
    & 35.1 & 1.6 & 0.80  \\[1 mm]
 S2-11 & $(1,2,-\frac{1}{2})_F$ & $(3,2,\frac{1}{6})_S$ &-& $1.2\cdot 10^{14}$ &
   38.4 & 1.2 & 0.61 \\[1 mm]
 S2-9 & $(3,1,-\frac{1}{3})_F$ & $(1,1,1)_S$ & $(1,3,0)_F$ (DM) & $2.6 \cdot 10^{14}$ & 37.9 & 0.54 & 0.28 \\[1 mm]
 T3-A, $m =-2$ & $(1,1,-1)_S$ & $(1,3,0)_S$ & $(1,2,-\frac{1}{2})_F$  & $4.0\cdot 10^{13}$ & 38.7 & 0.21 & 0.11 \\ [1 mm]
  \bottomrule
\end{tabular}
}
\caption{Collection of models that allow for gauge coupling unification. The name of the models, content of new particles Pi, scale of unification $\Lambda$ and value $\alpha^{-1} (\Lambda)$ of the gauge coupling at this scale are mentioned. Note that the first model in this table requires four generations of each new particle in order to lead to gauge coupling unification. In the last two columns, the relative errors on  the logarithm of $\Lambda$ and $\alpha^{-1} (\Lambda)$ are given according to the definitions in Eq.~(\ref{eq:delLambdaOverLambda}) in Sec.~\ref{sec:unification}.}\label{tab:conclusions}
\end{table}

In Sec.~\ref{sec:disc}, we estimate and exemplify the effect of the simplifications and approximations we have made in our analysis. 
In particular, we have studied the effect of the uncertainty in $\alpha_i$ at low energies, 
how the masses of the new particles affect our results, 
the importance of two-loop contributions to the RG running of the gauge couplings as well as 
 the impact of neglecting Yukawa couplings apart from the one of the top quark that is assumed to be
constant ($y_t=1$). The largest effects on the values of the unification scale $\Lambda$ and of the gauge coupling $\alpha^{-1}(\Lambda)$ are up to $3~\%$ 
 and originate from the variation of the masses of the new particles as well as from the two-loop contributions which, however, are always taken into account in the numerical analysis. 

In the present study, we ensure that in all models, which allow for gauge coupling unification, a possible LP in one of the gauge couplings occurs at a scale at least three orders of magnitude larger than 
the unification scale, see Eq.~(\ref{eq:LamLP}) in Sec.~\ref{sec:unification}.
Furthermore, LPs can be encountered in other couplings of the model, especially
in the quartic scalar couplings. In models with DM candidates in $\SU(2)$
representations with dimension larger than three, this might be an issue.
Following the literature, we have estimated in which models this can pose a
problem and found that this only happens in model S2-3 with the
DM candidate $(1,5,0)_S$.

Although the scale of unification is in most models lower than the one expected for a GUT, we have briefly discussed the possibility to embed the new particles in representations of $\SU(5)$. This is, indeed, 
possible for all particles with the largest necessary $\SU(5)$ representation being 200. 

An interesting possibility to explore is an intermediate scale where either new particles are introduced and/or the gauge symmetry is enhanced in order to raise the scale of gauge coupling unification.

\section*{Acknowledgments}

We would like to thank Stefan Antusch for useful discussions and Florian Lyonnet for very valuable help with the software PyR@TE.

This work was supported by the DFG Cluster of Excellence `Origin and Structure of the Universe' SEED project ``Neutrino mass generation mechanisms in (grand) unified flavor models and phenomenological imprints''. C.H.~and M.S.~would like to thank the Excellence Cluster `Universe', where part of this work was done. The CP$^3$-Origins center is partially funded by the Danish National Research Foundation, grant number DNRF90 (C.H.).
In addition, this work was supported in part by the Swedish Research Council (Vetenskapsr{\aa}det), contract no.~621-2011-3985 (T.O.~and S.R.) and the Australian Research Council (M.S.).  We acknowledge the use of \texttt{matplotlib}~\cite{Hunter:2007}, \texttt{ipython}~\cite{PER-GRA:2007} and \texttt{Julia}~\cite{2012arXiv1209.5145B,2014arXiv1411.1607B}.

\appendix

\mathversion{bold}
\section{One-Loop and Two-Loop Order Coefficients $b_k$ and $b_{k\ell}$}
\mathversion{normal}
\label{app:betacoeff}

In this appendix, we present the one- and two-loop order coefficients $b^i_k$ and $b^i_{k\ell}$, as defined in Eq.~\eqref{eq:bkbkl}, to the $\beta$-functions of the gauge couplings for scalars and fermions, contained in the models in Sec.~\ref{sec:models}. 
The coefficients $b_k^i$ for a complex scalar (CS) in the representation $(r_3,r_2,y)$ of the SM gauge group are given by
\begin{align}
\label{eq:b1CS}
	b_1^{\rm CS}&= \frac13 \, d(r_2)  \, d(r_3) \left( \frac35 y^2\right) \,,\\
	b_2^{\rm CS}&= \frac13 \, d(r_3) \, T(r_2) \,,\\
\label{eq:b3CS}
	b_3^{\rm CS}&= \frac13 \, d(r_2) \, T(r_3)
\end{align}
with $d(r)$ and $T(r)$ being the dimension and the Dynkin index of the representation $r$, respectively. The one-loop order coefficients $b_k^i$ for a real scalar, a Weyl fermion and a Dirac fermion can be obtained from the ones for a complex scalar by multiplying the coefficients $b_k^{\rm CS}$ with $\zeta=\frac12$, $2$ and $4$, respectively, i.e.
\begin{equation}
\label{eq:bscalarfermion}
	b_k^i = \zeta \, b_k^{\rm CS} = \begin{cases}
		b_k^{\rm CS} & \text{complex scalar}\\
		\frac12\, b_k^{\rm CS} & \text{real scalar}\\
		2\, b_k^{\rm CS} & \text{Weyl fermion}\\
		4\, b_k^{\rm CS} & \text{Dirac fermion}\\
	\end{cases} \,.
\end{equation}
Furthermore, note that the Dynkin index $T(r)$ of the representation $r$ is related to the eigenvalue $C_2(r)$ of the quadratic Casimir operator on $r$ 
\begin{equation}
	T(r) \, d(\mathrm{Adj}) = C_2(r) \, d(r) \,,
\end{equation}
where Adj denotes the adjoint representation. 
The dimension, the eigenvalue of the quadratic Casimir operator and the Dynkin index for representations $y$ of the hypercharge group are
\begin{equation} 
d(y)=1 \,, \quad T(y)=C_2(y) \quad \mbox{and} \quad C_2(y)= \frac 35 y^2 \,. 
\end{equation}
In particular, $C_2 (\mathrm{Adj})=0$ for the gauge group $\Uone$.
The quantities $d(r)$, $C_2(r)$ and $T(r)$ of an $\SU(2)$ representation $r$ as functions of the Dynkin label $\lambda$ (or the dimension $d(r)$) are given by
\begin{align}
	d(r)&=1+\lambda \,,\\
	C_2(r)&=\frac14\lambda(\lambda+2)= \frac14 \left(d(r)^2-1\right) \,,\\
	T(r)&=\frac{1}{12} \lambda\left(\lambda+1\right)\left(\lambda+2\right)=\frac{1}{12}\left(d(r)^3-d(r)\right) \,. \label{eq:dynkin}
\end{align}
Similarly, the dimension, the eigenvalue of the quadratic Casimir operator and the Dynkin index for an $\SU(3)$ representation $r$ with Dynkin labels $(\lambda_1, \lambda_2)$ are
\begin{align}
	d(r)&=\frac 12 \left(1+\lambda_1\right)\left(1+\lambda_2\right)\left(2+\lambda_1+\lambda_2\right) \,,\\
\label{eq:C2rSU3}
	C_2(r)&=\frac13
	\left(\lambda_1^2+\lambda_2^2+\lambda_1\lambda_2\right)+\lambda_1+\lambda_2 \,,\\
\label{eq:TrSU3}
	T(r)&=\frac{1}{48}\left(1+\lambda_1\right)\left(1+\lambda_2\right)\left(2+\lambda_1+\lambda_2\right)\left(\lambda_1^2+\lambda_2^2+\lambda_1\lambda_2+3 \lambda_1 +3 \lambda_2\right) \,.
\end{align}
The two-loop order coefficients $b_{k\ell}^i$ for a real scalar (Weyl fermion) are given by
\begin{equation}
\label{eq:bklscalarfermion}
	b_{k\ell}^i = \Delta c_{k\ell}^i + \xi\, \delta_{k\ell} \Delta c_{k}^i \,,
\end{equation}
where $\xi=1$ ($\xi=10$) for scalars (fermions) and the coefficients $\Delta c^i_{k\ell}$ and $\Delta c_k^i$ read 
\begin{align}	
\label{eq:Dckl}
	\Delta c_{k\ell}^i & =6 \,  b_k^i \, C_2(r_\ell) \,, \\
\label{eq:Dck}	
	\Delta c_{k}^i & =  b_k^i \, C_2(\mathrm{Adj}_k)
\end{align}
with $b_k^i$ being the one-loop order coefficients. Note that $\Delta c_1^i=0$, since the eigenvalue of the quadratic Casimir operator on the adjoint representation vanishes in the case of the gauge group $\Uone$.
The corresponding coefficients for a complex scalar (Dirac fermion) are obtained by multiplying the ones for a real scalar (Weyl fermion) by a factor of 2. In Tab.~\ref{tab:Bcoefficients}, we give the one-loop and two-loop order coefficients $b_k^i$
and $b_{k\ell}^i$ for a complex scalar and for various representations $(r_3,r_2,y)$ of the SM gauge group.
\newpage
\thispagestyle{empty}
\begin{table}[!htbp]
\centering
\begin{tabular}{cccc}
\toprule
Representation & One-loop & \multicolumn{2}{c}{$\phantom{xxxxxxx}$Two-loop}\\
$(r_3,r_2,y)$ & $ b_k^i$ & $\Delta c_{k\ell}^i$ & $\Delta c_{k}^i$ \\
\midrule
$(1,1,y)$&
$(\frac{1}{5} y^2,0,0)$&
$\begin{pmatrix}\frac{36}{25} y^4 & 0 & 0 \\0 & 0 & 0\\0 & 0 & 0\\\end{pmatrix}$&
$(0,0,0)$\\

$(1,2,y)$&
$(\frac{2}{5} y^2,\frac{1}{6},0)$&
$\begin{pmatrix}\frac{72}{25} y^4 & \frac{18}{5} y^2 & 0 \\[0.5mm]\frac{6}{5} y^2 & \frac{3}{2} & 0\\0 & 0 & 0\\\end{pmatrix}$&
$(0,\frac{2}{3},0)$\\

$(1,3,y)$&
$(\frac{3}{5} y^2,\frac{2}{3},0)$&
$\begin{pmatrix}\frac{108}{25} y^4 & \frac{72}{5} y^2 & 0 \\[0.5mm]\frac{24}{5} y^2 & 16 & 0\\0 & 0 & 0\\\end{pmatrix}$&
$(0,\frac{8}{3},0)$\\

$(1,4,y)$&
$(\frac{4}{5} y^2,\frac{5}{3},0)$&
$\begin{pmatrix}\frac{144}{25} y^4 & 36 y^2 & 0 \\12 y^2 & 75 & 0\\0 & 0 & 0\\\end{pmatrix}$&
$(0,\frac{20}{3},0)$\\

$(1,5,y)$&
$(y^2,\frac{10}{3},0)$&
$\begin{pmatrix}\frac{36}{5} y^4 & 72 y^2 & 0 \\24 y^2 & 240 & 0\\0 & 0 & 0\\\end{pmatrix}$&
$(0,\frac{40}{3},0)$\\

$(1,6,y)$&
$(\frac{6}{5} y^2,\frac{35}{6},0)$&
$\begin{pmatrix}\frac{216}{25} y^4 & 126 y^2 & 0 \\42 y^2 & \frac{1225}{2} & 0\\0 & 0 & 0\\\end{pmatrix}$&
$(0,\frac{70}{3},0)$\\

$(1,7,y)$&
$(\frac{7}{5} y^2,\frac{28}{3},0)$&
$\begin{pmatrix}\frac{252}{25} y^4 & \frac{1008}{5} y^2 & 0 \\[0.5mm]\frac{336}{5} y^2 & 1344 & 0\\0 & 0 & 0\\\end{pmatrix}$&
$(0,\frac{112}{3},0)$\\

$(3,1,y)$&
$(\frac{3}{5} y^2,0,\frac{1}{6})$&
$\begin{pmatrix}\frac{108}{25} y^4 & 0 & \frac{48}{5} y^2 \\0 & 0 & 0\\\frac{6}{5} y^2 & 0 & \frac{8}{3}\\\end{pmatrix}$&
$(0,0,1)$\\

$(3,2,y)$&
$(\frac{6}{5} y^2,\frac{1}{2},\frac{1}{3})$&
$\begin{pmatrix}\frac{216}{25} y^4 & \frac{54}{5} y^2 & \frac{96}{5} y^2 \\[0.5mm]\frac{18}{5} y^2 & \frac{9}{2} & 8\\[0.5mm]\frac{12}{5} y^2 & 3 & \frac{16}{3}\\\end{pmatrix}$&
$(0,2,2)$\\

$(3,3,y)$&
$(\frac{9}{5} y^2,2,\frac{1}{2})$&
$\begin{pmatrix}\frac{324}{25} y^4 & \frac{216}{5} y^2 & \frac{144}{5} y^2 \\[0.5mm]\frac{72}{5} y^2 & 48 & 32\\[0.5mm]\frac{18}{5} y^2 & 12 & 8\\\end{pmatrix}$&
$(0,8,3)$\\

$(8,1,y)$&
$(\frac{8}{5} y^2,0,1)$&
$\begin{pmatrix}\frac{288}{25} y^4 & 0 & \frac{288}{5} y^2 \\0 & 0 & 0\\\frac{36}{5} y^2 & 0 & 36\\\end{pmatrix}$&
$(0,0,6)$\\

$(8,2,y)$&
$(\frac{16}{5} y^2,\frac{4}{3},2)$&
$\begin{pmatrix}\frac{576}{25} y^4 & \frac{144}{5} y^2 & \frac{576}{5} y^2 \\[0.5mm]\frac{48}{5} y^2 & 12 & 48\\[0.5mm]\frac{72}{5} y^2 & 18 & 72\\\end{pmatrix}$&
$(0,\frac{16}{3},12)$\\

$(8,3,y)$&
$(\frac{24}{5} y^2,\frac{16}{3},3)$&
$\begin{pmatrix}\frac{864}{25} y^4 & \frac{576}{5} y^2 & \frac{864}{5} y^2 \\[0.5mm]\frac{192}{5} y^2 & 128 & 192\\[0.5mm]\frac{108}{5} y^2 & 72 & 108\\\end{pmatrix}$&
$(0,\frac{64}{3},18)$\\

\bottomrule
\end{tabular}
\caption{One-loop and two-loop order coefficients for different particles $(r_3,r_2,y)$. The values are given for a complex scalar. The coefficients for real scalars (where applicable) and fermions
can be derived by using the formulas in Eqs.~(\ref{eq:bscalarfermion}) and (\ref{eq:bklscalarfermion}).
Note the representations $(r_3,r_2,y)$ are ordered as $(\SU(3), \SU(2), \Uone)$, whereas the coefficients $b_k^i$, $\Delta c_{k\ell}^i$ and $\Delta c_k^i$ are ordered in the opposite way, i.e.~$(\Uone,\SU(2),\SU(3))$. 
}\label{tab:Bcoefficients}
\end{table}

\clearpage

\providecommand{\href}[2]{#2}\begingroup\raggedright\endgroup


\begin{thebibliography}{10}

\bibitem{Fukuda:1998mi}
{\scshape Super-Kamiokande} collaboration, Y.~Fukuda et~al., \emph{{Evidence
  for Oscillation of Atmospheric Neutrinos}},
  \href{http://dx.doi.org/10.1103/PhysRevLett.81.1562}{\emph{Phys. Rev. Lett.}
  {\bf 81} (1998) 1562--1567}, [\href{http://arxiv.org/abs/hep-ex/9807003}{{\tt
  hep-ex/9807003}}].

\bibitem{Ahmad:2002jz}
{\scshape SNO} collaboration, Q.~R. Ahmad et~al., \emph{{Direct Evidence for
  Neutrino Flavor Transformation from Neutral Current Interactions in the
  Sudbury Neutrino Observatory}},
  \href{http://dx.doi.org/10.1103/PhysRevLett.89.011301}{\emph{Phys. Rev.
  Lett.} {\bf 89} (2002) 011301},
  [\href{http://arxiv.org/abs/nucl-ex/0204008}{{\tt nucl-ex/0204008}}].

\bibitem{Lundmark:1930}
K.~Lundmark, \emph{{{\"U}ber die Bestimmung der Entfernungen, Dimensionen,
  Massen und Dichtigkeit f{\"u}r die n{\"a}chstgelegenen anagalaktischen
  Sternsysteme}}, {\emph{Medd. Lund. Obs.} {\bf 125} (1930) 1--13}.

\bibitem{Zwicky:1933xx}
F.~{Zwicky}, \emph{{Die Rotverschiebung von extragalaktischen Nebeln}},
  {\emph{Helv. Phys. Acta} {\bf 6} (1933) 110--127}.

\bibitem{Georgi:1974sy}
H.~Georgi and S.~L. Glashow, \emph{Unity of all elementary particle forces},
  \href{http://dx.doi.org/10.1103/PhysRevLett.32.438}{\emph{Phys. Rev. Lett.}
  {\bf 32} (1974) 438--441}.

\bibitem{Weinberg:1979sa}
S.~Weinberg, \emph{Baryon- and lepton-nonconserving processes},
  \href{http://dx.doi.org/10.1103/PhysRevLett.43.1566}{\emph{Phys. Rev. Lett.}
  {\bf 43} (1979) 1566--1570}.

\bibitem{Minkowski:1977sc}
P.~Minkowski, \emph{{$\mu \rightarrow e \gamma$} at a rate of one out of $10^9$
  muon decays?},
  \href{http://dx.doi.org/10.1016/0370-2693(77)90435-X}{\emph{Phys. Lett.} {\bf
  B67} (1977) 421--428}.

\bibitem{Yanagida:1980}
T.~Yanagida, \emph{Horizontal gauge symmetry and masses of neutrinos},  in
  \emph{Proceedings of the Workshop on The Unified Theory and the Baryon Number
  in the Universe} (O.~Sawada and A.~Sugamoto, eds.), p.~95, KEK, Tsukuba,
  Japan, 1979.

\bibitem{Glashow:1979vf}
S.~L. Glashow, \emph{The future of elementary particle physics},  in
  \emph{Proceedings of the 1979 Carg{\`e}se Summer Institute on Quarks and
  Leptons} (M.~L{\'e}vy, J.-L. Basdevant, D.~Speiser, J.~Weyers, R.~Gastmans
  and M.~Jacob, eds.), pp.~687--713, Plenum Press, New York, 1980.

\bibitem{Gell-Mann:1980vs}
M.~Gell-Mann, P.~Ramond and R.~Slansky, \emph{Complex spinors and unified
  theories},  in \emph{Supergravity} (P.~{van Nieuwenhuizen} and D.~Z.
  Freedman, eds.), p.~315, North Holland, Amsterdam, 1979.

\bibitem{Mohapatra:1979ia}
R.~N. Mohapatra and G.~Senjanovi{\'c}, \emph{{Neutrino Mass and Spontaneous
  Parity Nonconservation}},
  \href{http://dx.doi.org/10.1103/PhysRevLett.44.912}{\emph{Phys. Rev. Lett.}
  {\bf 44} (1980) 912--915}.

\bibitem{Magg:1980ut}
M.~Magg and C.~Wetterich, \emph{Neutrino mass problem and gauge hierarchy},
  \href{http://dx.doi.org/10.1016/0370-2693(80)90825-4}{\emph{Phys. Lett.} {\bf
  B94} (1980) 61--64}.

\bibitem{Schechter:1980gr}
J.~Schechter and J.~W.~F. Valle, \emph{{Neutrino masses in ${\rm SU(2)} \times
  {\rm U(1)}$ theories}},
  \href{http://dx.doi.org/10.1103/PhysRevD.22.2227}{\emph{Phys. Rev.} {\bf D22}
  (1980) 2227--2235}.

\bibitem{Wetterich:1981bx}
C.~Wetterich, \emph{{Neutrino Masses and the scale of $B-L$ violation}},
  \href{http://dx.doi.org/10.1016/0550-3213(81)90279-0}{\emph{Nucl. Phys.} {\bf
  B187} (1981) 343--375}.

\bibitem{Lazarides:1980nt}
G.~Lazarides, Q.~Shafi and C.~Wetterich, \emph{{Proton lifetime and fermion
  masses in an ${\rm SO(10)}$ model}},
  \href{http://dx.doi.org/10.1016/0550-3213(81)90354-0}{\emph{Nucl. Phys.} {\bf
  B181} (1981) 287--300}.

\bibitem{Mohapatra:1980yp}
R.~N. Mohapatra and G.~Senjanovi{\'c}, \emph{{Neutrino masses and mixings in
  gauge models with spontaneous parity violation}},
  \href{http://dx.doi.org/10.1103/PhysRevD.23.165}{\emph{Phys.Rev.} {\bf D23}
  (1981) 165--180}.

\bibitem{Cheng:1980qt}
T.~P. Cheng and L.-F. Li, \emph{{Neutrino masses, mixings and oscillations in
  ${\rm SU(2)} \times {\rm U(1)}$ models of electroweak interactions}},
  \href{http://dx.doi.org/10.1103/PhysRevD.22.2860}{\emph{Phys. Rev.} {\bf D22}
  (1980) 2860--2868}.

\bibitem{Foot:1988aq}
R.~Foot, H.~Lew, X.~G. He and G.~C. Joshi, \emph{Seesaw neutrino masses induced
  by a triplet of leptons},
  \href{http://dx.doi.org/10.1007/BF01415558}{\emph{Z. Phys.} {\bf C44} (1989)
  441--444}.

\bibitem{Zee:1980ai}
A.~Zee, \emph{A theory of lepton number violation and neutrino majorana
  masses}, \href{http://dx.doi.org/10.1016/0370-2693(80)90349-4}{\emph{Phys.
  Lett.} {\bf B93} (1980) 389--393}.

\bibitem{Zee:1985id}
A.~Zee, \emph{{Quantum numbers of Majorana neutrino masses}},
  \href{http://dx.doi.org/10.1016/0550-3213(86)90475-X}{\emph{Nucl. Phys.} {\bf
  B264} (1986) 99--110}.

\bibitem{Babu:1988ki}
K.~S. Babu, \emph{{Model of ``calculable'' Majorana neutrino masses}},
  \href{http://dx.doi.org/10.1016/0370-2693(88)91584-5}{\emph{Phys. Lett.} {\bf
  B203} (1988) 132--136}.

\bibitem{Krauss:2002px}
L.~M. Krauss, S.~Nasri and M.~Trodden, \emph{{Model for neutrino masses and
  dark matter}},
  \href{http://dx.doi.org/10.1103/PhysRevD.67.085002}{\emph{Phys. Rev.} {\bf
  D67} (2003) 085002}, [\href{http://arxiv.org/abs/hep-ph/0210389}{{\tt
  hep-ph/0210389}}].

\bibitem{Cheng:1977ir}
T.~P. Cheng and L.-F. Li, \emph{{Weak-interaction-induced neutrino
  oscillations}}, \href{http://dx.doi.org/10.1103/PhysRevD.17.2375}{\emph{Phys.
  Rev.} {\bf D17} (1978) 2375--2382}.

\bibitem{Dorsner:2005ii}
I.~Dorsner, P.~Fileviez~P{\'e}rez and R.~Gonz{\'a}lez~Felipe,
  \emph{{Phenomenological and cosmological aspects of a minimal GUT scenario}},
  \href{http://dx.doi.org/10.1016/j.nuclphysb.2006.05.006}{\emph{Nucl. Phys.}
  {\bf B747} (2006) 312--327}, [\href{http://arxiv.org/abs/hep-ph/0512068}{{\tt
  hep-ph/0512068}}].

\bibitem{Dorsner:2005fq}
I.~Dorsner and P.~Fileviez~P{\'e}rez, \emph{{Unification without supersymmetry:
  Neutrino mass, proton decay and light leptoquarks}},
  \href{http://dx.doi.org/10.1016/j.nuclphysb.2005.06.016}{\emph{Nucl. Phys.}
  {\bf B723} (2005) 53--76}, [\href{http://arxiv.org/abs/hep-ph/0504276}{{\tt
  hep-ph/0504276}}].

\bibitem{Arhrib:2009mz}
A.~Arhrib, B.~Bajc, D.~K. Ghosh, T.~Han, G.-Y. Huang, I.~Puljak et~al.,
  \emph{{Collider Signatures for Heavy Lepton Triplet in Type I+III Seesaw}},
  \href{http://dx.doi.org/10.1103/PhysRevD.82.053004}{\emph{Phys. Rev.} {\bf
  D82} (2010) 053004}, [\href{http://arxiv.org/abs/0904.2390}{{\tt
  0904.2390}}].

\bibitem{Dev:2015pga}
P.~S. Bhupal~Dev and R.~N. Mohapatra, \emph{{Unified Explanation of the $eejj$,
  Diboson, and Dijet Resonances at the LHC}},
  \href{http://dx.doi.org/10.1103/PhysRevLett.115.181803}{\emph{Phys. Rev.
  Lett.} {\bf 115} (2015) 181803}, [\href{http://arxiv.org/abs/1508.02277}{{\tt
  1508.02277}}].

\bibitem{Mohapatra:1986bd}
R.~N. Mohapatra and J.~W.~F. Valle, \emph{{Neutrino mass and baryon-number
  nonconservation in superstring models}},
  \href{http://dx.doi.org/10.1103/PhysRevD.34.1642}{\emph{Phys. Rev.} {\bf D34}
  (1986) 1642}.

\bibitem{Sayre:2006ma}
J.~Sayre, S.~Wiesenfeldt and S.~Willenbrock, \emph{{Minimal trinification}},
  \href{http://dx.doi.org/10.1103/PhysRevD.73.035013}{\emph{Phys. Rev.} {\bf
  D73} (2006) 035013}, [\href{http://arxiv.org/abs/hep-ph/0601040}{{\tt
  hep-ph/0601040}}].

\bibitem{Achiman:1978rv}
Y.~Achiman and B.~Stech, \emph{{Topless model for grand unification}},  in
  \emph{{Karlsruhe Summer Inst.1978:0303}}, p.~0303, 1978.

\bibitem{Glashow:1984gc}
S.~L. Glashow, \emph{{Trinification of all elementary particle forces}},  in
  \emph{{Providence Grand Unif.1984:0088}}, p.~0088, 1984.

\bibitem{Ma:2006km}
E.~Ma, \emph{{Verifiable radiative seesaw mechanism of neutrino mass and dark
  matter}}, \href{http://dx.doi.org/10.1103/PhysRevD.73.077301}{\emph{Phys.
  Rev.} {\bf D73} (2006) 077301},
  [\href{http://arxiv.org/abs/hep-ph/0601225}{{\tt hep-ph/0601225}}].

\bibitem{Parida:2011wh}
M.~K. Parida, \emph{{Radiative seesaw in ${\rm SO(10)}$ with dark matter}},
  \href{http://dx.doi.org/10.1016/j.physletb.2011.09.016}{\emph{Phys. Lett.}
  {\bf B704} (2011) 206--210}, [\href{http://arxiv.org/abs/1106.4137}{{\tt
  1106.4137}}].

\bibitem{Singer:1980sw}
M.~Singer, J.~W.~F. Valle and J.~Schechter, \emph{{Canonical neutral-current
  predictions from the weak-electromagnetic gauge group ${\rm SU(3)} \times
  {\rm U(1)}$}}, \href{http://dx.doi.org/10.1103/PhysRevD.22.738}{\emph{Phys.
  Rev.} {\bf D22} (1980) 738--743}.

\bibitem{Valle:1983dk}
J.~W.~F. Valle and M.~Singer, \emph{{Lepton-number violation with quasi-Dirac
  neutrinos}}, \href{http://dx.doi.org/10.1103/PhysRevD.28.540}{\emph{Phys.
  Rev.} {\bf D28} (1983) 540--545}.

\bibitem{Boucenna:2014dia}
S.~M. Boucenna, R.~M. Fonseca, F.~Gonz{\'a}lez-Canales and J.~W.~F. Valle,
  \emph{{Small neutrino masses and gauge coupling unification}},
  \href{http://dx.doi.org/10.1103/PhysRevD.91.031702}{\emph{Phys. Rev.} {\bf
  D91} (2015) 031702(R)}, [\href{http://arxiv.org/abs/1411.0566}{{\tt
  1411.0566}}].

\bibitem{Perez:2016qbo}
P.~Fileviez~P{\'e}rez and C.~Murgui, \emph{{Renormalizable ${\rm SU(5)}$
  Unification}},  \href{http://arxiv.org/abs/1604.03377}{{\tt 1604.03377}}.

\bibitem{Bouchand:2012dx}
R.~Bouchand and A.~Merle, \emph{{Running of radiative neutrino masses: the
  scotogenic model}},
  \href{http://dx.doi.org/10.1007/JHEP07(2012)084}{\emph{JHEP} {\bf 07} (2012)
  084}, [\href{http://arxiv.org/abs/1205.0008}{{\tt 1205.0008}}].

\bibitem{Babu:2014kca}
K.~S. Babu and J.~Julio, \emph{{Renormalization of a two-loop neutrino mass
  model}}, \href{http://dx.doi.org/10.1063/1.4883422}{\emph{AIP Conf. Proc.}
  {\bf 1604} (2014) 134--141}.

\bibitem{Cai:2014kra}
Y.~Cai, J.~D. Clarke, M.~A. Schmidt and R.~R. Volkas, \emph{{Testing radiative
  neutrino mass models at the LHC}},
  \href{http://dx.doi.org/10.1007/JHEP02(2015)161}{\emph{JHEP} {\bf 02} (2015)
  161}, [\href{http://arxiv.org/abs/1410.0689}{{\tt 1410.0689}}].

\bibitem{Babu:2001ex}
K.~S. Babu and C.~N. Leung, \emph{{Classification of effective neutrino mass
  operators}},
  \href{http://dx.doi.org/10.1016/S0550-3213(01)00504-1}{\emph{Nucl. Phys.}
  {\bf B619} (2001) 667--689}, [\href{http://arxiv.org/abs/hep-ph/0106054}{{\tt
  hep-ph/0106054}}].

\bibitem{deGouvea:2007xp}
A.~de~Gouv{\^e}a and J.~Jenkins, \emph{{Survey of lepton number violation via
  effective operators}},
  \href{http://dx.doi.org/10.1103/PhysRevD.77.013008}{\emph{Phys. Rev.} {\bf
  D77} (2008) 013008}, [\href{http://arxiv.org/abs/0708.1344}{{\tt
  0708.1344}}].

\bibitem{Angel:2012ug}
P.~W. Angel, N.~L. Rodd and R.~R. Volkas, \emph{{Origin of neutrino masses at
  the LHC: $\Delta L = 2$ effective operators and their ultraviolet
  completions}},
  \href{http://dx.doi.org/10.1103/PhysRevD.87.073007}{\emph{Phys. Rev. D} {\bf
  87} (2013) 073007}, [\href{http://arxiv.org/abs/1212.6111}{{\tt 1212.6111}}].

\bibitem{Restrepo:2013aga}
D.~Restrepo, O.~Zapata and C.~E. Yaguna, \emph{{Models with radiative neutrino
  masses and viable dark matter candidates}},
  \href{http://dx.doi.org/10.1007/JHEP11(2013)011}{\emph{JHEP} {\bf 11} (2013)
  011}, [\href{http://arxiv.org/abs/1308.3655}{{\tt 1308.3655}}].

\bibitem{Bonnet:2012kz}
F.~Bonnet, M.~Hirsch, T.~Ota and W.~Winter, \emph{{Systematic study of the
  $d=5$ Weinberg operator at one-loop order}},
  \href{http://dx.doi.org/10.1007/JHEP07(2012)153}{\emph{JHEP} {\bf 07} (2012)
  153}, [\href{http://arxiv.org/abs/1204.5862}{{\tt 1204.5862}}].

\bibitem{Angel:2013hla}
P.~W. Angel, Y.~Cai, N.~L. Rodd, M.~A. Schmidt and R.~R. Volkas,
  \emph{{Testable two-loop radiative neutrino mass model based on an
  $LLQd^cQd^c$ effective operator}},
  \href{http://dx.doi.org/10.1007/JHEP11(2014)092,
  10.1007/JHEP10(2013)118}{\emph{JHEP} {\bf 10} (2013) 118},
  [\href{http://arxiv.org/abs/1308.0463}{{\tt 1308.0463}}].

\bibitem{FileviezPerez:2009ud}
P.~Fileviez~P{\'e}rez and M.~B. Wise, \emph{{On the origin of neutrino
  masses}}, \href{http://dx.doi.org/10.1103/PhysRevD.80.053006}{\emph{Phys.
  Rev.} {\bf D80} (2009) 053006}, [\href{http://arxiv.org/abs/0906.2950}{{\tt
  0906.2950}}].

\bibitem{HALL198175}
L.~Hall, \emph{Grand unification of effective gauge theories},
  \href{http://dx.doi.org/10.1016/0550-3213(81)90498-3}{\emph{Nucl. Phys.} {\bf
  B178} (1981) 75--124}.

\bibitem{Machacek:1983tz}
M.~E. Machacek and M.~T. Vaughn, \emph{{Two-loop renormalization group
  equations in a general quantum field theory: (I). Wave function
  renormalization}},
  \href{http://dx.doi.org/10.1016/0550-3213(83)90610-7}{\emph{Nucl. Phys.} {\bf
  B222} (1983) 83--103}.

\bibitem{Deppisch:2016qqd}
F.~F. Deppisch, S.~Kulkarni, H.~P{\"a}s and E.~Schumacher, \emph{{Leptoquark
  patterns unifying neutrino masses, flavor anomalies, and the diphoton
  excess}}, \href{http://dx.doi.org/10.1103/PhysRevD.94.013003}{\emph{Phys.
  Rev.} {\bf D94} (2016) 013003}, [\href{http://arxiv.org/abs/1603.07672}{{\tt
  1603.07672}}].

\bibitem{Babu:2009aq}
K.~Babu, S.~Nandi and Z.~Tavartkiladze, \emph{{New mechanism for neutrino mass
  generation and triply charged Higgs bosons at the LHC}},
  \href{http://dx.doi.org/10.1103/PhysRevD.80.071702}{\emph{Phys. Rev.} {\bf
  D80} (2009) 071702(R)}, [\href{http://arxiv.org/abs/0905.2710}{{\tt
  0905.2710}}].

\bibitem{Babu:2010vp}
K.~Babu and J.~Julio, \emph{{Two-loop neutrino mass generation through
  leptoquarks}},
  \href{http://dx.doi.org/10.1016/j.nuclphysb.2010.07.022}{\emph{Nucl. Phys.}
  {\bf B841} (2010) 130--156}, [\href{http://arxiv.org/abs/1006.1092}{{\tt
  1006.1092}}].

\bibitem{Babu:2011vb}
K.~Babu and J.~Julio, \emph{{Radiative neutrino mass generation through
  vectorlike quarks}},
  \href{http://dx.doi.org/10.1103/PhysRevD.85.073005}{\emph{Phys. Rev.} {\bf
  D85} (2012) 073005}, [\href{http://arxiv.org/abs/1112.5452}{{\tt
  1112.5452}}].

\bibitem{He:2011hs}
X.-G. He and S.~K. Majee, \emph{{Implications of recent data on neutrino mixing
  and lepton flavour violating decays for the Zee model}},
  \href{http://dx.doi.org/10.1007/JHEP03(2012)023}{\emph{JHEP} {\bf 03} (2012)
  023}, [\href{http://arxiv.org/abs/1111.2293}{{\tt 1111.2293}}].

\bibitem{Cirelli:2005uq}
M.~Cirelli, N.~Fornengo and A.~Strumia, \emph{{Minimal dark matter}},
  \href{http://dx.doi.org/10.1016/j.nuclphysb.2006.07.012}{\emph{Nucl. Phys.}
  {\bf B753} (2006) 178--194}, [\href{http://arxiv.org/abs/hep-ph/0512090}{{\tt
  hep-ph/0512090}}].

\bibitem{Goodman:1984dc}
M.~W. Goodman and E.~Witten, \emph{{Detectability of certain dark-matter
  candidates}}, \href{http://dx.doi.org/10.1103/PhysRevD.31.3059}{\emph{Phys.
  Rev.} {\bf D31} (1985) 3059}.

\bibitem{Cirelli:2009uv}
M.~Cirelli and A.~Strumia, \emph{{Minimal dark matter: model and results}},
  \href{http://dx.doi.org/10.1088/1367-2630/11/10/105005}{\emph{New J. Phys.}
  {\bf 11} (2009) 105005}, [\href{http://arxiv.org/abs/0903.3381}{{\tt
  0903.3381}}].

\bibitem{Fischler:1981is}
M.~S. Fischler and C.~T. Hill, \emph{{Effects of large mass fermions on $M_X$
  and $\sin^2\theta_{\rm W}$}},
  \href{http://dx.doi.org/10.1016/0550-3213(81)90517-4}{\emph{Nucl. Phys.} {\bf
  B193} (1981) 53--60}.

\bibitem{Jones:1981we}
D.~R.~T. Jones, \emph{{Two-loop $\beta$-function for a $G_1 \times G_2$ gauge
  theory}}, \href{http://dx.doi.org/10.1103/PhysRevD.25.581}{\emph{Phys. Rev.}
  {\bf D25} (1982) 581--582}.

\bibitem{Agashe:2014kda}
{\scshape Particle Data Group} collaboration, K.~A. Olive et~al., \emph{{Review
  of Particle Physics}},
  \href{http://dx.doi.org/10.1088/1674-1137/38/9/090001}{\emph{Chin. Phys.}
  {\bf C38} (2014) 090001}.

\bibitem{Weinberg198051}
S.~Weinberg, \emph{Effective gauge theories},
  \href{http://dx.doi.org/10.1016/0370-2693(80)90660-7}{\emph{Phys. Lett.} {\bf
  B91} (1980) 51--55}.

\bibitem{Hisano:2003ec}
J.~Hisano, S.~Matsumoto and M.~M. Nojiri, \emph{{Explosive Dark Matter
  Annihilation}},
  \href{http://dx.doi.org/10.1103/PhysRevLett.92.031303}{\emph{Phys. Rev.
  Lett.} {\bf 92} (2004) 031303},
  [\href{http://arxiv.org/abs/hep-ph/0307216}{{\tt hep-ph/0307216}}].

\bibitem{Lyonnet:2013dna}
F.~Lyonnet, I.~Schienbein, F.~Staub and A.~Wingerter, \emph{{PyR@TE:
  Renormalization group equations for general gauge theories}},
  \href{http://dx.doi.org/10.1016/j.cpc.2013.12.002}{\emph{Comput. Phys.
  Commun.} {\bf 185} (2014) 1130--1152},
  [\href{http://arxiv.org/abs/1309.7030}{{\tt 1309.7030}}].

\bibitem{Buchmuller:1986zs}
W.~Buchm{\"u}ller, R.~R{\"u}ckl and D.~Wyler, \emph{{Leptoquarks in
  lepton-quark collisions}},
  \href{http://dx.doi.org/10.1016/0370-2693(87)90637-X}{\emph{Phys. Lett.} {\bf
  B191} (1987) 442--448}.

\bibitem{Dorsner:2016wpm}
I.~Dor{\v{s}}ner, S.~Fajfer, A.~Greljo, J.~F. Kamenik and N.~Ko{\v{s}}nik,
  \emph{{Physics of leptoquarks in precision experiments and at particle
  colliders}},
  \href{http://dx.doi.org/10.1016/j.physrep.2016.06.001}{\emph{Phys. Rep.} {\bf
  641} (2016) 1--68}, [\href{http://arxiv.org/abs/1603.04993}{{\tt
  1603.04993}}].

\bibitem{Pas:2015hca}
H.~P{\"a}s and E.~Schumacher, \emph{{Common origin of $R_K$ and neutrino
  masses}}, \href{http://dx.doi.org/10.1103/PhysRevD.92.114025}{\emph{Phys.
  Rev.} {\bf D92} (2015) 114025}, [\href{http://arxiv.org/abs/1510.08757}{{\tt
  1510.08757}}].

\bibitem{Sierra:2008wj}
D.~Aristizabal~Sierra, J.~Kubo, D.~Suematsu, D.~Restrepo and O.~Zapata,
  \emph{{Radiative seesaw model: Warm dark matter, collider signatures, and
  lepton flavor violating signals}},
  \href{http://dx.doi.org/10.1103/PhysRevD.79.013011}{\emph{Phys. Rev.} {\bf
  D79} (2009) 013011}, [\href{http://arxiv.org/abs/0808.3340}{{\tt
  0808.3340}}].

\bibitem{Farzan:2010mr}
Y.~Farzan, S.~Pascoli and M.~A. Schmidt, \emph{{AMEND: A model explaining
  neutrino masses and dark matter testable at the LHC and MEG}},
  \href{http://dx.doi.org/10.1007/JHEP10(2010)111}{\emph{JHEP} {\bf 10} (2010)
  111}, [\href{http://arxiv.org/abs/1005.5323}{{\tt 1005.5323}}].

\bibitem{Molinaro:2014lfa}
E.~Molinaro, C.~E. Yaguna and O.~Zapata, \emph{{FIMP realization of the
  scotogenic model}},
  \href{http://dx.doi.org/10.1088/1475-7516/2014/07/015}{\emph{JCAP} {\bf 07}
  (2014) 015}, [\href{http://arxiv.org/abs/1405.1259}{{\tt 1405.1259}}].

\bibitem{Vicente:2014wga}
A.~Vicente and C.~E. Yaguna, \emph{{Probing the scotogenic model with lepton
  flavor violating processes}},
  \href{http://dx.doi.org/10.1007/JHEP02(2015)144}{\emph{JHEP} {\bf 02} (2015)
  144}, [\href{http://arxiv.org/abs/1412.2545}{{\tt 1412.2545}}].

\bibitem{Restrepo:2015ura}
D.~Restrepo, A.~Rivera, M.~S{\'a}nchez-Pel{\'a}ez, O.~Zapata and W.~Tangarife,
  \emph{{Radiative neutrino masses in the singlet-doublet fermion dark matter
  model with scalar singlets}},
  \href{http://dx.doi.org/10.1103/PhysRevD.92.013005}{\emph{Phys. Rev.} {\bf
  D92} (2015) 013005}, [\href{http://arxiv.org/abs/1504.07892}{{\tt
  1504.07892}}].

\bibitem{Longas:2015sxk}
R.~Longas, D.~Portillo, D.~Restrepo and O.~Zapata, \emph{{The inert Zee
  model}}, \href{http://dx.doi.org/10.1007/JHEP03(2016)162}{\emph{JHEP} {\bf
  03} (2016) 162}, [\href{http://arxiv.org/abs/1511.01873}{{\tt 1511.01873}}].

\bibitem{Arhrib:2015dez}
A.~Arhrib, C.~B{\oe}hm, E.~Ma and T.-C. Yuan, \emph{{Radiative model of
  neutrino mass with neutrino interacting MeV dark matter}},
  \href{http://dx.doi.org/10.1088/1475-7516/2016/04/049}{\emph{JCAP} {\bf 04}
  (2016) 049}, [\href{http://arxiv.org/abs/1512.08796}{{\tt 1512.08796}}].

\bibitem{Ibarra:2016dlb}
A.~Ibarra, C.~E. Yaguna and O.~Zapata, \emph{{Direct detection of fermion dark
  matter in the radiative seesaw model}},
  \href{http://dx.doi.org/10.1103/PhysRevD.93.035012}{\emph{Phys. Rev.} {\bf
  D93} (2016) 035012}, [\href{http://arxiv.org/abs/1601.01163}{{\tt
  1601.01163}}].

\bibitem{vonderPahlen:2016cbw}
F.~von~der Pahlen, G.~Palacio, D.~Restrepo and O.~Zapata, \emph{{Radiative Type
  III Seesaw Model and its collider phenomenology}},
  \href{http://dx.doi.org/10.1103/PhysRevD.94.033005}{\emph{Phys. Rev.} {\bf
  D94} (2016) 033005}, [\href{http://arxiv.org/abs/1605.01129}{{\tt
  1605.01129}}].

\bibitem{Losada:2009yy}
M.~Losada and S.~Tulin, \emph{{Color Octet Leptogenesis}},
  \href{http://arxiv.org/abs/0909.0648}{{\tt 0909.0648}}.

\bibitem{FileviezPerez:2010ch}
P.~Fileviez~P{\'e}rez, T.~Han, S.~Spinner and M.~K. Trenkel, \emph{{Lepton
  Number Violation from Colored States at the LHC}},
  \href{http://dx.doi.org/10.1007/JHEP01(2011)046}{\emph{JHEP} {\bf 01} (2011)
  046}, [\href{http://arxiv.org/abs/1010.5802}{{\tt 1010.5802}}].

\bibitem{Choubey:2012ux}
S.~Choubey, M.~Duerr, M.~Mitra and W.~Rodejohann, \emph{{Lepton number and
  lepton flavor violation through color octet states}},
  \href{http://dx.doi.org/10.1007/JHEP05(2012)017}{\emph{JHEP} {\bf 05} (2012)
  017}, [\href{http://arxiv.org/abs/1201.3031}{{\tt 1201.3031}}].

\bibitem{Sierra:2014rxa}
D.~Aristizabal~Sierra, A.~Degee, L.~Dorame and M.~Hirsch, \emph{{Systematic
  classification of two-loop realizations of the Weinberg operator}},
  \href{http://dx.doi.org/10.1007/JHEP03(2015)040}{\emph{JHEP} {\bf 03} (2015)
  040}, [\href{http://arxiv.org/abs/1411.7038}{{\tt 1411.7038}}].

\bibitem{Ahriche:2014oda}
A.~Ahriche, K.~L. McDonald and S.~Nasri, \emph{{A model of radiative neutrino
  mass: with or without dark matter}},
  \href{http://dx.doi.org/10.1007/JHEP10(2014)167}{\emph{JHEP} {\bf 10} (2014)
  167}, [\href{http://arxiv.org/abs/1404.5917}{{\tt 1404.5917}}].

\bibitem{Ahriche:2014cda}
A.~Ahriche, C.-S. Chen, K.~L. McDonald and S.~Nasri, \emph{{Three-loop model of
  neutrino mass with dark matter}},
  \href{http://dx.doi.org/10.1103/PhysRevD.90.015024}{\emph{Phys. Rev.} {\bf
  D90} (2014) 015024}, [\href{http://arxiv.org/abs/1404.2696}{{\tt
  1404.2696}}].

\bibitem{Chen:2014ska}
C.-S. Chen, K.~L. McDonald and S.~Nasri, \emph{{A class of three-loop models
  with neutrino mass and dark matter}},
  \href{http://dx.doi.org/10.1016/j.physletb.2014.05.082}{\emph{Phys. Lett.}
  {\bf B734} (2014) 388--393}, [\href{http://arxiv.org/abs/1404.6033}{{\tt
  1404.6033}}].

\bibitem{Ahriche:2015wha}
A.~Ahriche, K.~L. McDonald, S.~Nasri and T.~Toma, \emph{{A model of neutrino
  mass and dark matter with an accidental symmetry}},
  \href{http://dx.doi.org/10.1016/j.physletb.2015.05.031}{\emph{Phys. Lett.}
  {\bf B746} (2015) 430--435}, [\href{http://arxiv.org/abs/1504.05755}{{\tt
  1504.05755}}].

\bibitem{Okada:2015hia}
H.~Okada and K.~Yagyu, \emph{{Three-loop neutrino mass model with doubly
  charged particles from isodoublets}},
  \href{http://dx.doi.org/10.1103/PhysRevD.93.013004}{\emph{Phys. Rev.} {\bf
  D93} (2016) 013004}, [\href{http://arxiv.org/abs/1508.01046}{{\tt
  1508.01046}}].

\bibitem{Ahriche:2015loa}
A.~Ahriche, K.~L. McDonald and S.~Nasri, \emph{{A Radiative Model for the Weak
  Scale and Neutrino Mass via Dark Matter}},
  \href{http://dx.doi.org/10.1007/JHEP02(2016)038}{\emph{JHEP} {\bf 02} (2016)
  038}, [\href{http://arxiv.org/abs/1508.02607}{{\tt 1508.02607}}].

\bibitem{Cai:2016jrl}
Y.~Cai and M.~A. Schmidt, \emph{{Revisiting the R$\nu$MDM models}},
  \href{http://dx.doi.org/10.1007/JHEP05(2016)028}{\emph{JHEP} {\bf 05} (2016)
  028}, [\href{http://arxiv.org/abs/1603.00255}{{\tt 1603.00255}}].

\bibitem{Ahriche:2016rgf}
A.~Ahriche, K.~L. McDonald, S.~Nasri and I.~Picek, \emph{{A critical analysis
  of one-loop neutrino mass models with minimal dark matter}},
  \href{http://dx.doi.org/10.1016/j.physletb.2016.04.022}{\emph{Phys. Lett.}
  {\bf B757} (2016) 399--404}, [\href{http://arxiv.org/abs/1603.01247}{{\tt
  1603.01247}}].

\bibitem{Hamada:2015bra}
Y.~Hamada, K.~Kawana and K.~Tsumura, \emph{{Landau pole in the Standard Model
  with weakly interacting scalar fields}},
  \href{http://dx.doi.org/10.1016/j.physletb.2015.05.072}{\emph{Phys. Lett.}
  {\bf B747} (2015) 238--244}, [\href{http://arxiv.org/abs/1505.01721}{{\tt
  1505.01721}}].

\bibitem{Cai:2015kpa}
C.~Cai, Z.-M. Huang, Z.~Kang, Z.-H. Yu and H.-H. Zhang, \emph{{Perturbativity
  limits for scalar minimal dark matter with Yukawa Interactions: Septuplet}},
  \href{http://dx.doi.org/10.1103/PhysRevD.92.115004}{\emph{Phys. Rev.} {\bf
  D92} (2015) 115004}, [\href{http://arxiv.org/abs/1510.01559}{{\tt
  1510.01559}}].

\bibitem{Slansky:1981yr}
R.~Slansky, \emph{{Group theory for unified model building}},
  \href{http://dx.doi.org/10.1016/0370-1573(81)90092-2}{\emph{Phys. Rep.} {\bf
  79} (1981) 1--128}.

\bibitem{Patera:1981pn}
J.~Patera and R.~T. Sharp, \emph{{${\rm SU(3)} \times {\rm SU(2)} \times {\rm
  U(1)}$ content of all ${\rm SU(5)}$ representations}},
  \href{http://dx.doi.org/10.1103/PhysRevD.25.1141}{\emph{Phys. Rev.} {\bf D25}
  (1982) 1141--1142}.

\bibitem{Hunter:2007}
J.~D. Hunter, \emph{Matplotlib: A {2D} {G}raphics {E}nvironment},
  \href{http://dx.doi.org/10.1109/MCSE.2007.55}{\emph{Comput. Sci. Eng.} {\bf
  9} (2007) 90--95}.

\bibitem{PER-GRA:2007}
F.~P\'erez and B.~E. Granger, \emph{{IP}ython: A {S}ystem for {I}nteractive
  {S}cientific {C}omputing},
  \href{http://dx.doi.org/10.1109/MCSE.2007.53}{\emph{Comput. Sci. Eng.} {\bf
  9} (2007) 21--29}.

\bibitem{2012arXiv1209.5145B}
J.~{Bezanson}, S.~{Karpinski}, V.~B. {Shah} and A.~{Edelman}, \emph{{Julia: A
  Fast Dynamic Language for Technical Computing}},
  \href{http://arxiv.org/abs/1209.5145}{{\tt 1209.5145}}.

\bibitem{2014arXiv1411.1607B}
J.~{Bezanson}, A.~{Edelman}, S.~{Karpinski} and V.~B. {Shah}, \emph{{Julia: A
  Fresh Approach to Numerical Computing}},
  \href{http://arxiv.org/abs/1411.1607}{{\tt 1411.1607}}.

\end{thebibliography}


\end{document}